\documentclass[12pt,aps]{revtex4}
\usepackage{epsfig}

\newcommand{\beeq}{\begin{equation}}
\newcommand{\eneq}{\end{equation}}
\newcommand{\be}{\begin{eqnarray}}
\newcommand{\ee}{\end{eqnarray}}
\newcommand{\bpic}{\begin{picture}}
\newcommand{\epic}{\end{picture}}
\newcommand{\bs}{\begin{scriptsize}}
\newcommand{\es}{\end{scriptsize}}

\def\dd{\partial}
\def\la{\raise.16ex\hbox{$\langle$} \, }
\def\ra{\, \raise.16ex\hbox{$\rangle$} }

\def\d{\delta}
\def\e{\epsilon}
\def\g{\gamma}
\def\G{\Gamma}
\def\r{\rho}

\def\D{\Delta}

\def\l{\lambda}

\def\Box{\kern1pt\vbox{\hrule height 1.2pt\hbox{\vrule width 1.2pt\hskip 3pt
   \vbox{\vskip 6pt}\hskip 3pt\vrule width 0.6pt}\hrule height 0.6pt}\kern1pt}
\def\gtwid{\mathrel{\raise.3ex\hbox{$>$\kern-.75em\lower1ex\hbox{$\sim$}}}}
\def\ltwid{\mathrel{\raise.3ex\hbox{$<$\kern-.75em\lower1ex\hbox{$\sim$}}}}
\def\Fint{\rlap{$\Biggl\rfloor$}\Biggl\lceil}
\def\Box{\kern1pt\vbox{\hrule height 1.2pt\hbox{\vrule width 1.2pt\hskip 3pt
   \vbox{\vskip 6pt}\hskip 3pt\vrule width 0.6pt}\hrule height 0.6pt}\kern1pt}
\begin{document}
\begin{titlepage}
\begin{flushright}
gr-qc/0408080\\LPT-ORSAY-04-64 \\ UFIFT-QG-04-5 \\t04/103\\
\end{flushright}
\vspace{.4cm}
\begin{center}
\textbf{Two Loop Scalar Self-Mass during Inflation}
\end{center}
\begin{center}
T. Brunier$^*$
\end{center}
\begin{center}
\textit{Service de Physique Th\'eorique, CEA/Saclay, 91191 Gif-sur-Yvette
FRANCE}
\end{center}
\begin{center}
V. K. Onemli$^{\dagger}$
\end{center}
\begin{center}
\textit{Laboratoire de Physique Th\'eorique, Universit\'e Paris
XI, B\^at. 210, 91405 Orsay FRANCE}
\end{center}
\begin{center}
R. P. Woodard$^{\ddagger}$
\end{center}
\begin{center}
\textit{Department of Physics, University of Florida, Gainesville, FL 32611
USA}
\end{center}

\begin{center}
ABSTRACT
\end{center}
We work in the locally de Sitter background of an inflating universe and
consider a massless, minimally coupled scalar with a quartic self-interaction.
We use dimensional regularization to compute the fully renormalized scalar
self-mass-squared at one and two loop order for a state which is released in
Bunch-Davies vacuum at $t=0$. Although the field strength and coupling
constant renormalizations are identical to those of flat space, the geometry
induces a non-zero mass renormalization. The finite part also shows a sort of
growing mass that competes with the classical force in eventually turning
off this system's super-acceleration.

\begin{flushleft}
\end{flushleft}
\vspace{.4cm}
\begin{flushleft}
$^*$ e-mail: brunier@spht.saclay.cea.fr \\
$^{\dagger}$ e-mail: Vakif.Onemli@th.u-psud.fr \\
$^{\ddagger}$ e-mail: woodard@phys.ufl.edu
\end{flushleft}
\end{titlepage}

\section{Introduction}

For studying quantum fields which are spectators during inflation
it is convenient to model the background geometry using a portion
of the full de Sitter manifold known as the open coordinate patch.
If the $D$-dimensional cosmological constant is $\Lambda \equiv
(D\!-\!1) H^2$, the invariant element can be expressed, in either
conformal or co-moving coordinates, as follows:
\begin{equation}
ds^2 = a^2 \Bigl(-d\eta^2 + d\vec{x} \! \cdot \! d\vec{x}\Bigr)
= -dt^2 + a^2 d\vec{x} \! \cdot \! d\vec{x} \qquad {\rm where}
\qquad a(\eta) = -\frac1{H \eta} = e^{Ht} \; . \label{ds^2}
\end{equation}
One might expect that the inflationary expansion of this spacetime
makes quantum effects stronger by allowing virtual particles to
persist longer than in flat space. Indeed, it is simple to see
that any sufficiently long wavelength (${\rm wavelength}>1/H$)
virtual particle which is massless on the Hubble scale can exist
forever \cite{RPW}. However, one must also consider the {\it rate}
at which virtual particles emerge from the vacuum. Classical
conformal invariance causes this rate (per unit co-moving time) to
fall like $1/a$, so any long wave length virtual particles which
emerge become real, but very few emerge \cite{RPW}. To get
enhanced quantum effects during inflation therefore requires
quanta which are effectively massless and also not classically
conformally invariant.

A fascinating example of enhanced quantum effects occurs when a massless
minimally coupled scalar is endowed with a quartic self-interaction,
\begin{equation}
\mathcal{L} = -\frac12 \partial_{\mu} \varphi \partial_{\nu} \varphi
g^{\mu\nu} \sqrt{-g} - \frac{\lambda}{4!} \varphi^4 \sqrt{-g} + {\rm
counterterms} \; . \label{simL}
\end{equation}
If this system is released at $t=0$ ($\eta = -1/H$) in free Bunch-Davies
vacuum the fully renormalized energy density and pressure are \cite{OW1,OW2},
\begin{eqnarray}
\r_{\mbox{\tiny{ren}}} & = & \frac{\Lambda}{8\pi G} + \frac{\l H^4}{2^6\pi^4}
\Bigg\{ \frac12 \ln^2\left(a\right) + \frac29 a^{-3} - \frac12
\sum_{n=1}^{\infty}
\frac{n+2}{(n+1)^2} a^{-n-1} \Bigg\} + O(\lambda^2) \; , \label{rho} \\
p_{\mbox{\tiny{ren}}} & = & - \frac{\Lambda}{8\pi G} - \frac{\l H^4}{2^6\pi^4}
\Bigg\{\frac12 \ln^2\left(a\right) + \frac13 \ln\left(a\right) +
\frac16 \sum_{n=1}^{\infty} \frac{n^2-4}{(n+1)^2}
a^{-n-1}\Bigg\} + O(\lambda^2) \; . \label{pres}
\end{eqnarray}
Hence their sum violates the weak energy condition on cosmological scales,
\be
\rho_{\mbox{\tiny{ren}}} + p_{\mbox{\tiny{ren}}} =
\frac{\l H^4}{2^6\pi^4} \Bigg\{- \frac13 \ln\left(a\right) +
\frac29 a^{-3} - \frac16 \sum_{n=1}^{\infty} \frac{n+2}{n+1}
a^{-n-1}\Bigg\} + O(\lambda^2) \; . \label{it}
\ee

The physics behind this surprising result is that inflationary particle
production causes the scalar to undergo a random walk such that its average
distance from the minimum of the $\varphi^4$ potential increases without
bound in the free theory \cite{VF,ADL,AAS},
\begin{equation}
\Bigl\langle \Omega \Bigl\vert \varphi^2(x) \Bigr\vert \Omega \Bigr\rangle_0
= \Bigl({\rm Divergent\ Constant}\Bigr) + \frac{H^2}{4\pi^2} \ln(a) \; .
\end{equation}
After the ultraviolet divergence is removed by renormalization the effect
is to drive the scalar up its potential, which increases the vacuum energy
and leads to (\ref{rho}) and (\ref{pres}). The process is self-limiting for
two reasons. First, as the scalar rises up its potential the classical
$\varphi^3$ force pushes it back down. Second, the $\varphi^2$ curvature
associated with being away from $\varphi = 0$ acts like a positive mass-squared
to reduce the inflationary particle production responsible for pushing the
scalar away from $\varphi = 0$.

This model is of more than passing interest because the supernova
acceleration data are consistent with an evolving dark energy
equation of state whose current value is less than minus one
\cite{ASSS,CP,WM,G,NP,WT,ASS,FWZ,JBP,CKPCB,HM,VETTBFT}. Such a
thing is difficult to explain with classical models on account of
the problem with stability
\cite{CHT,CJM,NOO,NO1,NO2,NO3,Bar0,NO4,NO5,Sriv0,ENO,NO6,LA,MMMA,
AP,BM,FT,Wei0,HJW,Sriv,V,BDE,Noj,FLPZ,AGP,CS,CFT,HN,CMS,BP,YHW1,YHW2,LS,WMeng,Bar,Sriv2,
TR,FLHZ,BdVS,RAW,CKT,SLe,CL00,RL,CL0,
ACL,CL,GC,MM,SNLP,RJS,MSA,Hu,GPZZ,PPA,TPad,CW,Huey,Eas,Gong,Ded,XFZ,ZKG,Sriv3,CGST}.
Therefore one is lead naturally to a self-limiting quantum effect
in a classically stable theory such as (\ref{simL}).

Although this model gives a value for $w +1$ which is far too small to explain
the observations \cite{OW2}, it does serve to establish that quantum effects
can induce a temporary phase of super-acceleration. Further, it does so in a
wonderfully simple setting for which every issue can be settled by explicit
computation. To check stability one first computes the self-mass-squared
$M^2(x;x')$ and then studies how it modifies the effective field equations
at linearized order,
\begin{equation}
\Box \varphi(x) - \int d^4x' M^2(x;x') \varphi(x') = 0 \; . \label{linord}
\end{equation}
That first step is the object of this paper. Just as in flat space one
must go to two loop order to see corrections to the derivative terms, so we
will compute the fully renormalized self-mass-squared at one and two loop
orders. In a later work we will employ the result to study the behavior of
linearized perturbations in the quantum-corrected field equations.

In Section II we explain the Schwinger-Keldysh formalism that one must use
for cosmological settings in which the state is specified at a finite time
rather than in the asymptotic past and future. We also derive the Feynman
rules. The one loop counterterms are computed in Section III. Sections IV--VI
treat the various two loop contributions to the self-mass-squared. In Section
VII we give the fully renormalized results. Implications are discussed in
Section VIII. Section IX is an appendix devoted to various technical results
used throughout.

\section{Feynman Rules in the Schwinger-Keldysh Formalism}

The Schwinger-Keldysh formalism is a technique that makes computing
expectation values almost as simple as the Feynman rules do for computing
in-out matrix elements \cite{JS,RJ}. To sketch the derivation, consider
a real scalar field, $\phi(x)$ whose Lagrangian (not Lagrangian density)
at time $t$ is $L[\phi(t)]$. Most quantum field theorists are familiar
with the functional integral expression for the matrix element of an
operator $\mathcal{O}_1[\phi]$ between states whose wave functionals are
given at a starting time $s$ and a last time $\ell$,
\begin{equation}
\Bigl\langle \Phi \Bigl\vert T^*\Bigl(\mathcal{O}_1[\phi]\Bigr) \Bigr\vert
\Psi \Bigr\rangle = \Fint [d\phi] \, \mathcal{O}_1[\phi] \,
\Phi^*[\phi(\ell)] \, e^{i \int_{s}^{\ell} dt L[\phi(t)]} \,
\Psi[\phi(s)] \; . \label{1stint}
\end{equation}
The $T^*$-ordering symbol in the matrix element indicates that the operator
$\mathcal{O}_1[\phi]$ is time-ordered, except that any derivatives are
taken {\it outside} the time-ordering. We can use (\ref{1stint}) to obtain a
similar expression for the matrix element of the {\it anti}-time-ordered
product of some operator $\mathcal{O}_2[\phi]$ in the presence of the
reversed states,
\begin{eqnarray}
\Bigl\langle \Psi \Bigl\vert \overline{T}^*\Bigl(\mathcal{O}_2[\phi]\Bigr)
\Bigr\vert \Phi \Bigr\rangle & = & \Bigl\langle \Phi \Bigl\vert T^*\Bigl(
\mathcal{O}_2^{\dagger}[\phi]\Bigr) \Bigr\vert \Psi \Bigr\rangle^* \\
& = & \Fint [d\phi] \, \mathcal{O}_2[\phi] \, \Phi[\phi(\ell)] \,
e^{-i \int_{s}^{\ell} dt L[\phi(t)]} \, \Psi^*[\phi(s)] \; .
\label{2ndint}
\end{eqnarray}

Now note that summing over a complete set of states $\Phi$ gives a delta
functional,
\begin{equation}
\sum_{\Phi} \Phi\Bigl[\phi_-(\ell)\Bigr] \, \Phi^*\Bigl[\phi_+(\ell)
\Bigr] = \delta\Bigl[\phi_-(\ell) \!-\! \phi_+(\ell) \Bigr]
\; . \label{sum}
\end{equation}
Taking the product of (\ref{1stint}) and (\ref{2ndint}), and using (\ref{sum}),
we obtain a functional integral expression for the expectation value of any
anti-time-ordered operator $\mathcal{O}_2$ multiplied by any time-ordered
operator $\mathcal{O}_1$,
\begin{eqnarray}
\lefteqn{\Bigl\langle \Psi \Bigl\vert \overline{T}^*\Bigl(\mathcal{O}_2[
\phi]\Bigr) T^*\Bigl(\mathcal{O}_1[\phi]\Bigr) \Bigr\vert \Psi
\Bigr\rangle = \Fint [d\phi_+] [d\phi_-] \, \delta\Bigl[\phi_-(\ell)
\!-\! \phi_+(\ell)\Bigr] } \nonumber \\
& & \hspace{1.5cm} \times \mathcal{O}_2[\phi_-] \mathcal{O}_1[\phi_+]
\Psi^*[\phi_-(s)] e^{i \int_s^{\ell} dt \Bigl\{L[\phi_+(t)] -
L[\phi_-(t)]\Bigr\}} \Psi[\phi_+(s)] \; . \qquad \label{fund}
\end{eqnarray}
This is the fundamental relation between the canonical operator formalism
and the functional integral formalism in the Schwinger-Keldysh formalism.

The Feynman rules follow from (\ref{fund}) in close analogy to those for
in-out matrix elements. Because the same field is represented by two different
dummy functional variables, $\phi_{\pm}(x)$, the endpoints of lines carry
a ${\scriptscriptstyle \pm}$ polarity. External lines associated with the
operator $\mathcal{O}_2[\phi]$ have ${\scriptscriptstyle -}$ polarity whereas
those associated with the operator $\mathcal{O}_1[\phi]$ have
${\scriptscriptstyle +}$ polarity. Interaction vertices are either all
${\scriptscriptstyle +}$ or all ${\scriptscriptstyle -}$. Vertices with
${\scriptscriptstyle +}$ polarity are the same as in the usual
Feynman rules whereas vertices with the ${\scriptscriptstyle -}$ polarity
have an additional minus sign. Propagators can be ${\scriptscriptstyle ++}$,
${\scriptscriptstyle -+}$, ${\scriptscriptstyle +-}$ and ${\scriptscriptstyle
--}$.

The four propagators can be read off from the fundamental relation (\ref{fund})
when the free Lagrangian is substituted for the full one. It is useful to
denote canonical expectation values in the free theory with a subscript $0$.
With this convention we see that the ${\scriptscriptstyle ++}$ propagator is
just the ordinary Feynman propagator,
\begin{equation}
i\Delta_{\scriptscriptstyle ++}(x;x') = \Bigl\langle \Omega \Bigl\vert
T\Bigl(\phi(x) \phi(x') \Bigr) \Bigr\vert \Omega \Bigr\rangle_0 =
i\Delta(x;x') \; . \label{++}
\end{equation}
The other cases are simple to read off and to relate to the Feynman propagator,
\begin{eqnarray}
i\Delta_{\scriptscriptstyle -+}(x;x') \!\!\! & = & \!\!\! \Bigl\langle \Omega
\Bigl\vert \phi(x) \phi(x') \Bigr\vert \Omega \Bigr\rangle_0 =
\theta(t\!-\!t') i\Delta(x;x') \!+\! \theta(t'\!-\!t) \Bigl[i\Delta(x;x')
\Bigr]^* \! , \quad \label{-+} \\
i\Delta_{\scriptscriptstyle +-}(x;x') \!\!\! & = & \!\!\! \Bigl\langle \Omega
\Bigl\vert \phi(x') \phi(x) \Bigr\vert \Omega \Bigr\rangle_0 =
\theta(t\!-\!t') \Bigl[i\Delta(x;x')\Bigr]^* \!\!+\! \theta(t'\!-\!t)
i\Delta(x;x') , \quad \label{+-} \\
i\Delta_{\scriptscriptstyle --}(x;x') \!\!\! & = & \!\!\! \Bigl\langle \Omega
\Bigl\vert \overline{T}\Bigl(\phi(x) \phi(x') \Bigr) \Bigr\vert
\Omega \Bigr\rangle_0 = \Bigl[i\Delta(x;x')\Bigr]^* . \label{--}
\end{eqnarray}
Therefore we can get the four propagators of the Schwinger-Keldysh formalism
from the Feynman propagator once that is known.

Quantum field theory has much the same structure in the Schwinger-Keldysh
formalism as for the familiar in-out formalism. In particular,
one-particle-irreducible (1PI) diagrams mean the same thing and play the same
fundamental role. Because each external line can take either polarity there
are $2^N$ variations for each $N$-point function of the in-out formalism.
They generate an effective action that depends upon background fields
$\phi_+(x)$ and $\phi_-(x)$. For example, there are four Schwinger-Keldysh
2-point 1PI functions, $M^2_{\scriptscriptstyle \pm\pm}(x;x')$. The
${\scriptscriptstyle ++}$ one is the same as the in-out self-mass-squared
and the others are related as the propagators,
\begin{equation}
-i M^2_{\scriptscriptstyle --}(x;x') = \Bigl(-i M^{2}_{\scriptscriptstyle ++}(
x;x') \Bigr)^* \qquad , \qquad -i M^2_{\scriptscriptstyle -+}(x;x') = \Bigl(
-i M^2_{\scriptscriptstyle +-}(x;x')\Bigr)^* .
\end{equation}
The various self-mass-squared's enter the effective action as follows \cite{RJ},
\begin{eqnarray}
\lefteqn{\Gamma[\phi_+,\phi_-] = S[\phi_+] - S[\phi_-]} \nonumber \\
& & - \frac12 \int d^Dx \int d^Dx' \left\{ \matrix{ \phi_+(x) M^2_{
\scriptscriptstyle ++}(x;x') \phi_+(x') + \phi_+(x) M^2_{\scriptscriptstyle
+-}(x;x') \phi_-(x') \cr +\phi_-(x) M^2_{\scriptscriptstyle -+}(x;x')
\phi_+(x') + \phi_-(x) M^2_{\scriptscriptstyle --}(x;x') \phi_-(x') }
\right\} + O(\phi^3_{\pm}) . \qquad
\end{eqnarray}
Note that since counterterms derive from the classical actions --- $S[\phi_+]$
and $-S[\phi_-]$ --- there can be no primitive divergences in 1PI diagrams
of mixed polarity.

The effective field equations of the Schwinger-Keldysh formalism are obtained
by varying with respect to either background field and then setting the two
fields equal \cite{RJ},
\begin{equation}
\frac{\delta \Gamma[\phi_+,\phi_-]}{\delta \phi_+(x)} \Biggl\vert_{\phi_{\pm} =
\phi} = \frac{\delta S[\phi]}{\delta \phi(x)} - \int d^Dx' \Bigl\{
M^2_{\scriptscriptstyle ++}(x;x') + M^2_{\scriptscriptstyle +-}(x;x') \Bigr\}
\phi(x') + O(\phi^2) = 0 \; .
\end{equation}
Just as in-out matrix elements of the field obey the in-out effective field
equations, so expectation values of the field obey the Schwinger-Keldysh
effective field equations. Unlike the in-out effective field equations,
the Schwinger-Keldysh effective field equations are causal and purely real.
That is, although the equations are generally not local, there is no net
dependence at point $x^{\mu}$ from fields at any point $x^{\prime \mu}$ which
lies outside the past light-cone of $x^{\mu}$. Hence the initial value
problem is well defined. Also, if the initial value data are real, the
solution remains real.

Up to this point the discussion has been general. We begin the process of
deriving the Feynman rules of our model by expressing its Lagrangian
(\ref{simL}) in terms of the unrenormalized field $\phi$ and bare parameters
$m_0^2$ and $\lambda_0$,
\begin{equation}
\mathcal{L} = -\frac12 \partial_{\mu} \phi \partial_{\nu} \phi g^{\mu \nu}
\sqrt{-g} - \frac12 m_0^2 \phi^2 \sqrt{-g} - \frac1{4!} \lambda_0 \phi^4
\sqrt{-g} \; .
\end{equation}
Specializing to the (D-dimensional) locally de Sitter geometry in conformal
coordinates (\ref{ds^2}), and introducing the renormalized field $\varphi
\equiv \phi/\sqrt{Z}$ gives,
\begin{equation}
\mathcal{L} = -\frac12 Z \partial_{\mu} \varphi \partial_{\nu} \varphi
\eta^{\mu \nu} a^{D-2} - \frac12 Z m_0^2 \varphi^2 a^D - \frac1{4!} Z^2
\lambda_0 \phi^4 a^D \; .
\end{equation}
We now make the usual decomposition of bare parameters into renormalized
parameters and counterterms,
\begin{equation}
Z \equiv 1 + {\delta Z} \qquad , \qquad Z m_0^2 \equiv 0 + {\delta m}^2
\qquad , \qquad Z^2 \lambda_0 \equiv \lambda + {\delta \lambda} \; .
\end{equation}
Note that we define the renormalized mass to be zero. Of course this can
only be enforced initially; the point of the calculation is to see what
happens later under the impact of inflationary particle production!

The final expression of the Lagrangian distinguishes the free part (1st
term), the interaction (2nd term) and the counter-interactions (last 3
terms),
\begin{equation}
\mathcal{L} = -\frac12 \partial_{\mu} \varphi \partial_{\nu} \varphi
\eta^{\mu \nu} a^{D-2} -\frac1{4!} \lambda \varphi^4 a^D
-\frac12 {\delta Z} \partial_{\mu} \varphi \partial_{\nu} \varphi
\eta^{\mu \nu} a^{D-2} - \frac12 {\delta m}^2 \varphi^2 a^D - \frac1{4!}
{\delta \lambda} \phi^4 a^D \; .
\end{equation}
The propagator derives from the free part and obeys the equation,
\begin{equation}
{\rm Propagator} \qquad \Longrightarrow \qquad \eta^{\mu\nu} \partial_{\mu}
\Bigl(a^{D-2} \partial_{\nu} i \Delta(x;x')\Bigr) = i \delta^D(x-x') \; .
\label{propeqn}
\end{equation}
The position-space interaction vertex depends upon the four points $x^{\mu}_{
1-4}$,
\begin{equation}
{\rm Interaction} \qquad \Longrightarrow \qquad -i \lambda a_1^D
\delta^D(x_1 \!-\! x_2) \delta^D(x_2 \!-\! x_3) \delta^D(x_3 \!-\! x_4)
\; . \label{4pt}
\end{equation}
The field strength renormalization depends upon two points $x^{\mu}$ and
$x^{\prime \mu}$,
\begin{equation}
{\rm Field\ Strength\ Renormalization} \qquad \Longrightarrow \qquad
i {\delta Z} \eta^{\mu\nu} \partial_{\mu} \Bigl(a^{D-2} \partial_{\nu}
\delta^D(x \!-\! x') \Bigr) \; .
\end{equation}
The mass renormalization also depends upon $x^{\mu}$ and $x^{\prime \mu}$,
\begin{equation}
{\rm Mass\ Renormalization} \qquad \Longrightarrow \qquad -i {\delta m}^2 a^D
\delta^D(x \!-\! x') \; .
\end{equation}
Of course the coupling constant renormalization is the same as (\ref{4pt})
with $\lambda$ replaced by $\delta \lambda$,
\begin{equation}
{\rm Coupling\ Constant\ Renorm.} \quad \Longrightarrow \quad
-i {\delta \lambda} a_1^D \delta^D(x_1 \!-\! x_2) \delta^D(x_2 \!-\! x_3)
\delta^D(x_3 \!-\! x_4) \; .
\end{equation}

It remains to solve (\ref{propeqn}) for the propagator. One might think
to look for a function of the de Sitter invariant geodesic length
$\ell(x;x')$ but there is no such solution. The reason is that the free
theory possesses no normalizable, de Sitter invariant wave functions, and
hence no de Sitter invariant states \cite{AF}. By allowing $i\Delta(x;x')$
to depend additionally upon the scale factors $a \equiv a(\eta)$ and $a'
\equiv a(\eta')$ one can find a solution which has the homogeneity and
isotropy of cosmology \cite{OW1}. It is expressed in terms of a length
function we call $y_{\scriptscriptstyle ++}(x;x')$,
\begin{equation}
y_{\scriptscriptstyle ++}(x;x') \equiv a a' H^2 \left[ \Vert \vec{x} \!-\!
\vec{x}' \Vert^2 \!-\! (\vert \eta \!-\! \eta' \vert \!-\! i \d)^2 \right]
\equiv a a' H^2 {\Delta x}^2_{\scriptscriptstyle ++}(x;x') \; ,\\
\end{equation}
When the infinitesimal parameter $\delta$ vanishes this has a simple
relation to the geodesic length, $y = 4 \sin^2(H \ell/2)$. One can see from
(\ref{++}-\ref{--}) that the four propagators must all depend the same way
on $a$, $a'$ and length functions $y_{\scriptscriptstyle \pm\pm}$ defined as
follows,
\begin{eqnarray}
y_{\scriptscriptstyle +-}(x;x') & \equiv & a a' H^2 \left[\Vert \vec{x} \!-\!
\vec{x}' \Vert^2 \!-\! (\eta \!-\! \eta' \!+\! i \d)^2 \right] \equiv a a' H^2
{\Delta x}^2_{\scriptscriptstyle +-}(x;x') \; ,\\
y_{\scriptscriptstyle -+}(x;x') & \equiv & a a' H^2 \left[\Vert \vec{x} \!-\!
\vec{x}' \Vert^2 \!-\! (\eta \!-\! \eta' \!-\! i \d)^2 \right] \equiv a a' H^2
{\Delta x}^2_{\scriptscriptstyle -+}(x;x') \; ,\\
y_{\scriptscriptstyle --}(x;x') & \equiv & a a' H^2 \left[\Vert \vec{x} \!-\!
\vec{x}' \Vert^2 \!-\! (\vert \eta \!-\! \eta' \vert \!+\! i \d)^2 \right]
\equiv a a' H^2 {\Delta x}^2_{\scriptscriptstyle --}(x;x') \; .
\end{eqnarray}
Because of this close relationship between the four propagators we shall
often drop the ${\scriptscriptstyle \pm}$ subscripts and treat expressions
for general $y=y(x;x')$ and ${\Delta x}^2 = {\Delta x}^2(x;x')$.

To facilitate dimensional regularization we express the dimension of spacetime
in terms of its deviation from four, $D = 4 - \epsilon$. It is also convenient
to break the propagator up into three terms,
\begin{eqnarray}
i\Delta(x;x') & = & A(x;x') + B(x;x') + C(x;x') \; , \label{prop}\\
A(x;x') & \equiv & \frac{\Gamma(1 \!-\! \frac{\epsilon}2)}{4 \pi^{2 \!-\!
\frac{\epsilon}2}} \frac{(a a')^{-1 + \frac{\epsilon}2}}{\Delta x^{2 -
\epsilon}} \; , \label{split} \\
B(x;x') & \equiv & \frac{H^{2 - \epsilon}}{(4 \pi)^{2 - \frac{\epsilon}2}}
\left\{- \frac{2 \Gamma(3 \!-\! \frac{\epsilon}2)}{\epsilon} \Bigl(\frac{y}4
\Bigr)^{\frac{\epsilon}2} + \pi {\rm cot}\Bigl(\frac{\pi \epsilon}2\Bigr)
\frac{\Gamma(3 \!-\! \epsilon)}{\Gamma(2 \!-\! \frac{\epsilon}2)} +
\frac{\Gamma(3 \!-\! \epsilon)}{\Gamma(2 \!-\! \frac{\epsilon}2)} \ln(a a')
\right\} , \\
C(x;x') & \equiv & \frac{H^{2 - \epsilon}}{(4 \pi)^{2 - \frac{\epsilon}2}}
\sum_{n=1}^{\infty} \left[\frac1{n} \frac{\Gamma(3 \!-\! \epsilon \!+\! n)}{
\Gamma(2 \!-\! \frac{\epsilon}2 \!+\! n)} \Bigl(\frac{y}4\Bigr)^n -
\frac1{n \!+\! \frac{\epsilon}2} \frac{\Gamma(3 \!-\! \frac{\epsilon}2 \!+\!
n)}{ \Gamma(2 \!+\! n)} \Bigl(\frac{y}4\Bigr)^{n + \frac{\epsilon}2}\right] .
\end{eqnarray}
Powers of $A(x;x')$ are the ultimate source of ultraviolet divergences. The
term $B(x;x')$ is not ultraviolet divergent by itself,
\begin{equation}
\lim_{\epsilon \rightarrow 0} B(x;x') = \Bigl(\frac{H}{4 \pi}\Bigr)^2
\left\{ -2 \ln(H^2 {\Delta x}^2/4) - 1 \right\} = -\frac{H^2}{8
\pi^2} \ln\Bigl( \frac{\sqrt{e}}4 H^2 {\Delta x}^2 \Bigr) .
\end{equation}
However, we must retain $\epsilon \neq 0$ whenever $B(x;x')$ multiplies two
or more powers of $A(x;x')$. The term $C(x;x')$ actually vanishes for
$\epsilon = 0$. We only need it when multiplying by more than two
powers of $A(x;x')$, which can not happen for the maximum number of
propagators --- $(i\D(x;x'))^3$ --- we shall encounter in this computation.

\section{One Loop Counterterms}

Neither the non-flat geometry, nor the Schwinger-Keldysh formalism, nor even
the use of position space rather than momentum space, alters the basic
structure of perturbative quantum field theory. In particular, the diagrams
all look the same as in flat momentum space, and working at two loops requires
knowledge of the one loop counterterms. The purpose of this section is to
compute the order $\lambda$ contribution to $\delta m^2$ and the order
$\lambda^2$ contribution to $\delta \lambda$. As in flat space there is no
one loop contribution to $\delta Z$.

\begin{figure}[htbp]
\centerline{\epsfig{file=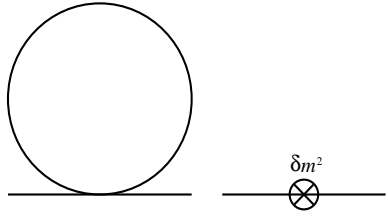,
                               width=2.0in,height=1.1in }}
\caption{Generic one loop self-mass-squared. \label{fig1} }
\end{figure}

Fig.~\ref{fig1} depicts the one loop contribution to the scalar
self-mass-squared. We can extract $\delta m^2$ at order $\lambda$
by focusing on the ${\scriptscriptstyle ++}$ term, \be -i
M^2_{\scriptscriptstyle 1++}(x;x')= -i \left[\frac{\l}2 i\D_{
\scriptscriptstyle ++}(x;x) + \d m^2 \right]a^D \d^D(x \!-\! x') +
O(\lambda^2) \; . \ee Although the coincidence limit of the
massless propagator vanishes when using dimensional regularization
in flat space, the de Sitter curvature of our background gives a
nonzero result that can be read off from (\ref{prop}), \be
i\D(x';x') = \lim_{x\rightarrow x'} i\D(x;x') =
\frac{H^{2-\e}}{(4\pi)^{2-\frac{\e}{2}}} \frac{\G(3 \!-\!
\e)}{\G(2 \!-\! \frac{\e}{2})} \Bigg\{2\ln(a) +
\pi\cot\Bigl(\frac{\pi\e}{2}\Bigr)\Bigg\} \; .\label{coin} \ee
Because of the finite, time-dependent term we cannot make the one
loop self-mass-squared vanish for all time. Our renormalization
condition is that it should be zero at $t=0$, which implies, \be
\d m^2 = -\frac{\l H^{2-\e}}{2^{5-\e}\pi^{2-{\frac{\e}2}}}
\frac{\G\left(3 \!-\! \e\right)}{\G\left(2\!-\! \frac{\e}2\right)}
\pi \cot\Bigl(\frac{\pi\e}{2}\Bigr) + O(\l^2) \; . \label{massct}
\ee The renormalized ${\scriptscriptstyle ++}$ self-mass-squared
is therefore, \be M^2_{\scriptscriptstyle 1++}(x;x') = \frac{\l
H^2}{8\pi^2} \,a^4 \ln(a) \delta^4(x-x') + O(\l^2) \; .
\label{M1++} \ee The ${\scriptscriptstyle --}$ case is minus the
complex conjugate,
\begin{equation}
M^2_{\scriptscriptstyle 1--}(x;x') = - \frac{\l H^2}{8\pi^2} \,a^4
\ln(a) \delta^4(x-x') + O(\l^2) \; .
\end{equation}
The ${\scriptscriptstyle +-}$ and ${\scriptscriptstyle -+}$ cases vanish at
this order because there are no mixed interactions,
\begin{equation}
M^2_{\scriptscriptstyle 1+-}(x;x') =
-\Bigl(M^2_{\scriptscriptstyle 1-+}(x;x') \Bigr)^* = 0 +
O(\lambda^2) \; .
\end{equation}

\begin{figure}[htbp]
\centerline{\epsfig{file=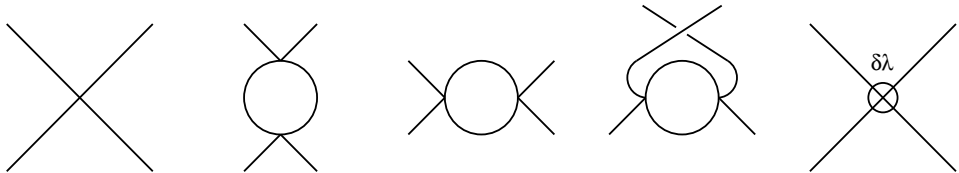,
                               width=5.0in,height=0.9in }}
\caption{Generic tree and one loop vertex function. \label{fig2} }
\end{figure}

We can extract the order $\lambda^2$ contribution to $\delta \lambda$ from
the one loop corrections to the ${\scriptscriptstyle ++++}$ 4-point 1PI
function. Fig.~\ref{fig2} shows the diagram topology. A simple application
of the Feynman rules gives,
\begin{eqnarray}
\lefteqn{-iV_+(x_1,x_2,x_3,x_4) = -i (\lambda \!+\! {\delta \lambda}) a_1^D
\delta^D(x_1 \!-\! x_2) \delta^D(x_1 \!-\! x_3) \delta^D(x_1 \!-\! x_4) -
\frac12 \lambda^2 \Biggl\{ } \nonumber \\
& & a_1^D a_4^D \Bigl[i \Delta_{\scriptscriptstyle ++}(x_1;x_4)\Bigr]^2
\delta^D(x_1 \!-\! x_2) \delta^D(x_3 \!-\! x_4) \!+\! a_1^D a_3^D \Bigl[i
\Delta_{\scriptscriptstyle ++}(x_1;x_3)\Bigr]^2 \delta^D(x_1 \!-\! x_4)
\delta^D(x_2 \!-\! x_3) \nonumber \\
& & \hspace{5cm} + a_1^D a_2^D \Bigl[i\Delta_{\scriptscriptstyle ++}(x_1;x_2)
\Bigr]^2 \delta^D(x_1 \!-\! x_3) \delta^D(x_4 \!-\! x_2) \Biggr\}
+ O(\lambda^3) \; . \qquad
\end{eqnarray}
In Appendix B we show that the square of a propagator can be expressed as a
local term which diverges for $\epsilon \longrightarrow 0$, plus a nonlocal
term which is integrable for $\epsilon = 0$. For the ${\scriptscriptstyle ++}$
propagator the result is equation (\ref{D2++}) which we reproduce here,
\begin{eqnarray}
\Bigl(i \Delta_{\scriptscriptstyle ++}(x;x') \Bigr)^2 & \longrightarrow &
- \frac{i \mu^{-\epsilon}}{2^3 \pi^{2 \!-\! \frac{\epsilon}2}} \frac{\Gamma(1
\!-\! \frac{\epsilon}2)}{(1 \!-\! \epsilon) \epsilon} \, a^{-4 \!+\! 2\epsilon}
\delta^D(x\!-\! x') - \frac{(a a')^{-2}}{2^6 \pi^4} \partial^2 \left(
\frac{\ln(\mu^2 {\Delta x}_{\scriptscriptstyle ++}^2)}{{\Delta x}_{
\scriptscriptstyle ++}^2} \right) \nonumber \\
& & \hspace{2.5cm} - \frac{H^2}{2^4 \pi^4} \frac{\ln\Bigl( \frac{\sqrt{e}}4 H^2
{\Delta x}^2_{\scriptscriptstyle ++} \Bigr)}{a a' {\Delta x}^2_{
\scriptscriptstyle ++}} + \frac{H^4}{2^6 \pi^4} \ln^2\Bigl( \frac{\sqrt{e}}4
H^2 {\Delta x}^2_{\scriptscriptstyle ++} \Bigr) . \qquad
\end{eqnarray}
To fix $\delta \lambda$ we are interested only in the local, divergent
term. There is such a contribution from the $s$-channel ---
$\Bigl(i \Delta_{\scriptscriptstyle ++}(x_1;x_4) \Bigr)^2$ --- the $t$-channel
--- $\Bigl(i \Delta_{\scriptscriptstyle ++}(x_1;x_3) \Bigr)^2$ --- and the
$u$-channel --- $\Bigl(i \Delta_{\scriptscriptstyle ++}(x_1;x_2) \Bigr)^2$ ---
graphs. They each make the same contribution, so the divergent part of the
vertex is,
\begin{eqnarray}
\lefteqn{-iV_+(x_1,x_2,x_3,x_4) = -i (\lambda + {\delta \lambda})
a_1^{4- \epsilon} \delta^D(x_1 - x_2) \delta^D(x_1 - x_3)
\delta^D(x_1 - x_4) }
\nonumber \\
& & \hspace{2cm} + \frac{i 3 \lambda^2\mu^{-\e}}{2^4 \pi^{2 \!-\!
\frac{\epsilon}2}} \frac{\Gamma(1 \!-\! \frac{\epsilon}2)}{(1
\!-\! \epsilon) \epsilon} \, a^4_1 \delta^D(x_1 - x_2)
\delta^D(x_1 - x_3) \delta^D(x_1 - x_4)\nonumber\\&& \hspace{6cm}+
\lambda^2 \times \Bigl({\rm Nonlocal\ Finite}\Bigr) + O(\lambda^3)
\; .
\end{eqnarray}
The simplest renormalization condition is to absorb the entire
local, constant part at order $\lambda^2$,
\be
\d\l=\frac{3\l^2 \mu^{-\e}}{16\pi^{2-\frac{\e}{2}}}
\frac{\G(1-\frac{\e}{2})}{\e(1-\e)} \; . \label{dl}
\ee
Note that while the divergence is proportional to the fourth power of the
scale factor, the counterterm is proportional to the $D$-th power. This
means that expanding $a_1^{4-\e}=a_1^4(1-\e \ln(a_1)+O(\e^2))$, gives an
additional, finite local term. The full renormalized ${\scriptscriptstyle
++++}$ 1PI 4-point vertex is,
\begin{eqnarray}
&&-iV_{\rm R}(x_1,x_2,x_3,x_4) \!=\! -i \left(\lambda -\frac{3\l^2}{2^4 \pi^2}
\ln(a_1)\right) a_1^4 \delta^4(x_1 \!-\! x_2)
\delta^4(x_1 \!-\! x_3) \delta^4(x_1 \!-\! x_4)\!+\!
\frac{\lambda^2}{2^7 \pi^4} \!\Bigg\{ \nonumber\\&&(a_1 a_4)^2
\partial^2_1 \! \left[ \frac{\ln(\mu^2
{\Delta x}^2_{14})}{\Delta x_{14}^2} \right] + 4 H^2 (a_1 a_4)^3
\frac{\ln\Bigl( \frac{\sqrt{e}}{4} H^2 {\Delta x}_{14}^2
\Bigr)}{\Delta x_{14}^2}-H^4 (a_1 a_4)^4
\ln^2\Bigl(\frac{\sqrt{e}}4 H^2 {\Delta x}_{14}^2 \Bigr)
\Bigg\}\nonumber\\&&\hspace{2cm} \times\delta^4(x_1 \!-\! x_2)
\delta^4(x_3 \!-\! x_4)+ \lambda^2 \Bigl(t {\rm -channel}\Bigr) +
\lambda^2 \Bigl(u {\rm -channel} \Bigr) + O (\lambda^3) \; .
\end{eqnarray}

\section{The Sunset Diagram}

\begin{figure}[htbp]
\centerline{\epsfig{file=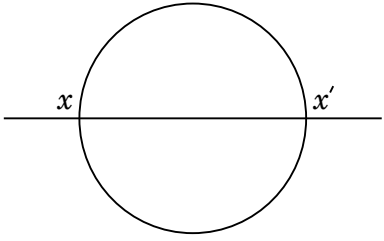,
                               width=2.0in,height=1.22in }}
\caption{Generic Sunset Diagram. \label{fig3} }
\end{figure}

The first two loop contribution to the scalar self-mass-squared is known as
the sunset (ss) diagram. Fig.~\ref{fig3} depicts its topology. A simple
application of the Feynman rules gives,
\begin{equation}
-i M^2_{\scriptscriptstyle {\rm ss} \pm\pm} = \frac1{3!} (\mp i\lambda)
(\mp i \lambda) (aa')^{4-\epsilon} \Bigl(i\D_{\scriptscriptstyle \pm\pm}(x;x')
\Bigr)^3 \; .
\end{equation}

In Appendix C we show that the cube of a propagator can be expressed as a
local term which diverges for $\epsilon \longrightarrow 0$, plus a nonlocal
term which is integrable for $\epsilon = 0$. For the ${\scriptscriptstyle ++}$
propagator the result is equation (\ref{D3++}) which we reproduce here,
\be
&&(i\D_{++}(x ;
x'))^3\rightarrow -\frac{i\mu^{-2\e} \Gamma^2(1-\frac{\e}2)}{2^9\;
\pi^{4-\e}} \; \frac{(a a')^{-3+ \frac{3}{2}\e}
\partial^2 \delta^D(x-x')}{ (1 \!-\! \frac{3}{2}\e ) (1 \!-\! \e)(1
\!-\!\frac{3}{4} \e)
\e}+\frac{i3H^{2-\e}\mu^{-\e}a^{-4+2\e}}{2^{6-\e} \pi^{4-\e}}
\;\Bigg\{ \nonumber \\
&& \frac{(1 - \frac{\epsilon}2)(1-\frac{\e}{4})\G^2(1-\frac{\e}{2})}{(1-\frac32
\e)\e^2}\; \Bigl(\frac{H a}{2\mu}\Bigr)^{\epsilon}-\frac{\G(1-\e)}{\e}\left[\pi
\cot\left(\frac{\pi \epsilon}{2}\right)+2\ln{(a)}\right]
\Bigg\}\d^D(x-x') \nonumber \\
&& - \frac{(a a')^{-3}}{2^{11} \pi^6} \,
\partial^4 \! \left[ \frac{\ln(\mu^2 \Delta x^2_{\scriptscriptstyle ++})}{
\Delta x^2_{\scriptscriptstyle ++}} \right]+
\frac{3H^2(aa')^{-2}}{2^8\pi^6}
\partial^2\Bigg[\ln\Big(\frac{He^{\frac{3}{4}}}{2\mu}\Big)
\frac{\ln(\mu^2 \Delta x^2_{++})}{\Delta
x^2_{++}}+\frac{1}{4}\frac{\ln^2(\mu^2 \Delta x^2_{++})}{\Delta
x^2_{++}}\Bigg]\nonumber\\&&\hspace{5.5cm}+\frac{3 H^4}{2^8 \pi^6}
\frac{\ln^2\Bigl( \frac{\sqrt{e}}4 H^2 \Delta x_{++}^2 \Bigr)}{a
a' {\Delta x^2_{++}}} - \frac{H^6}{2^9 \pi^6} \ln^3\Bigl(
\frac{\sqrt{e}}4 H^2 \Delta x^2_{++} \Bigr)\; . \ee When this is
multiplied by $(aa')^{4-\e}$ a few simple rearrangements allow us
to recognize the spacetime dependent part of the the first term as
essentially a field strength renormalization,
\begin{eqnarray}
(aa')^{1 + \frac{\epsilon}2} \partial^2 \delta^D(x \!-\! x') & = &
a^{1 + \frac{\epsilon}2} \partial^2 \Bigl( a^{1 + \frac{\epsilon}2}
\delta^D(x \!-\! x') \Bigr) \nonumber \\
& = & a^{2 + \epsilon} \left\{ \partial^2 - (2 \!+\! \epsilon) H a \partial_0
- \Bigl(1 \!+\! \frac{\epsilon}2\Bigr) \Bigl(2 + \frac{\epsilon}2\Bigr) H^2 a^2
\right\} \delta^D(x \!-\! x') \; . \label{massre}
\end{eqnarray}
Doing the same thing for the second term reveals what is essentially a mass
renormalization plus some finite local terms {\it and} an overlapping
divergence,
\begin{eqnarray}
\lefteqn{\frac{i3 H^2 a^4}{2^6 \pi^4}
\Bigl(\frac{2\pi}{H\mu}\Bigr)^{ \epsilon} \! \Bigg\{ \!\! \frac{(1
\!-\! \frac{\epsilon}4) (1 \!-\! \frac{\epsilon}{2}) \Gamma^2(1
\!-\! \frac{\epsilon}{2})}{\epsilon^2(1 \!-\! \frac32 \epsilon)}
\Bigl(\frac{H a}{2\mu}\Bigr)^{\epsilon} \!\!-\! \frac{ \Gamma(1
\!-\! \epsilon)}{\epsilon} \! \left[\pi \cot\Bigl(\frac{\pi
\epsilon}{2}\Bigr) \!+\! 2 \ln{(a)} \right] \!\! \Bigg\}
\delta^D(x \!-\! x')}
\nonumber \\
& &\longrightarrow \frac{i 3 H^2 a^4}{2^6 \pi^4}
\Bigl(\frac{2\pi}{H\mu}\Bigr)^{\epsilon} \! \Bigg\{ \!\! \frac{(1
\!-\! \frac{\epsilon}4) (1 \!-\! \frac{\epsilon}{2}) \Gamma^2(1
\!-\! \frac{\epsilon}{2})}{\epsilon^2(1 \!-\! \frac32 \epsilon)}
\Bigl(\frac{H}{2\mu}\Bigr)^{\epsilon} \!\!-\! \frac{\Gamma(1 \!-\!
\epsilon)}{
\epsilon} \pi \cot\Bigl(\frac{\pi \epsilon}{2}\Bigr) \nonumber \\
& & \hspace{4cm} - \frac1{\epsilon} \ln{(a)} \!+\! \frac12
\ln^2(a) \!+\! \left[\frac34 \!-\!
\gamma+\ln\left(\frac{H}{2\mu}\right)\right] \ln(a) \Bigg\}
\delta^D(x \!-\! x') \; . \qquad \label{over}
\end{eqnarray}

In displaying the final result we exhibit the field strength and mass
renormalizations, as well as the overlapping divergence and the finite,
local terms. Our result for the ${\scriptscriptstyle ++}$ case is,
\be
& & M^2_{\mbox{{\tiny ss$++$}}} \longrightarrow - \frac{\l^2}{2^{10} 3\pi^4}
\Bigl(\frac{\pi}{\mu^2} \Bigr)^\e \frac{\G^2(1-\frac{\e}2)}{(1 \!-\!
\frac34\e) (1 \!-\! \frac32\e) (1 \!-\!  \e) \e} \, a^2 \Bigl[\partial^2
\!-\! 2 H a \partial_0 \!-\! 2 H^2 a^2\Bigr] \delta^D(x \!-\! x') \nonumber \\
& & - \frac{\lambda^2}{2^{10} 3\pi^4} \, a^2 \Biggl\{ \ln(a) \partial^2 \!-\!
\Bigl(2 \ln(a) \!+\! 1\Bigr) H a \partial_0 \!-\! \Bigl(2 \ln(a) \!+\!
\frac32\Bigr) H^2 a^2 \Biggr\} \delta^4(x \!-\! x') \nonumber \\
& & + \frac{\l^2 H^2}{2^7 \pi^4} \Bigl( \frac{2 \pi}{H \mu} \Bigr)^\e
\Bigg\{ \frac{(1 \!-\! \frac{\epsilon}4) (1 \!-\! \frac{\e}2) \G^2(1 \!-\!
\frac{\e}2)}{\e^2 (1 \!-\! \frac32 \e)} \Bigl(\frac{H}{2\mu}\Bigr)^{\epsilon}
\!-\! \frac{\G(1 \!-\! \e)}{\e} \pi \cot\left(\frac{\pi \epsilon}{2}\right)
\Biggr\} a^4 \delta^D(x \!-\! x') \nonumber \\
& & - \frac{\lambda^2 H^2}{2^7 \pi^4} \Bigl(\frac{2 \pi}{H
\mu}\Bigr)^{ \epsilon} \, \frac{\ln(a)}{\epsilon} \, a^4
\delta^D(x \!-\! x') \!+\! \frac{ \lambda^2 H^2}{2^7 \pi^4}
\Biggl\{ \frac12 \ln^2(a) \!+\! \left[\frac34 \!-\!
\gamma+\ln\left(\frac{H}{2\mu}\right)\right] \ln(a)\Biggr\} a^4 \delta^4(x \!-\! x') \nonumber \\
& & + \frac{i \l^2 a a'}{2^9 \pi^6} \Bigg\{\frac1{24} \dd^4
\Biggl[ \frac{\ln\left(\mu^2 \D x^2_{\scriptscriptstyle
++}\right)}{\D x^2_{ \scriptscriptstyle ++}} \Biggr] \!-\! H^2 a
a' \dd^2 \Bigg[ \ln\Bigl( \frac{He^{\frac34}}{2 \mu} \Bigr)
\frac{\ln\left(\mu^2 \D x^2_{ \scriptscriptstyle ++}\right)}{\D
x^2_{\scriptscriptstyle ++}} \!+\! \frac{\ln^2\left(\mu^2 \D
x^2_{\scriptscriptstyle ++} \right)}{4 \D
x^2_{\scriptscriptstyle ++}} \Bigg] \nonumber \\
& & \hspace{3.5cm} - H^4 (a a')^2 \frac{\ln^2\left(\frac{\sqrt{e}}4 H^2
\D x^2_{\scriptscriptstyle ++}\right)}{\D x^2_{\scriptscriptstyle ++}} \!+\!
\frac{H^6}6 (a a')^3 \ln^3\Bigl(\frac{\sqrt{e}}4 H^2 \D x^2_{\scriptscriptstyle
++}\Bigr) \Bigg\}\; . \label{Mss++}
\ee
The ${\scriptscriptstyle --}$ case is minus the complex conjugate,
\be
& & M^2_{\mbox{{\tiny ss$--$}}} \longrightarrow \frac{\l^2}{2^{10} 3\pi^4}
\Bigl(\frac{\pi}{\mu^2} \Bigr)^\e \frac{\G^2(1-\frac{\e}2)}{(1 \!-\!
\frac34\e) (1 \!-\! \frac32\e) (1 \!-\!  \e) \e} \, a^2 \Bigl[\partial^2
\!-\! 2 H a \partial_0 \!-\! 2 H^2 a^2\Bigr] \delta^D(x \!-\! x') \nonumber \\
& & + \frac{\lambda^2}{2^{10} 3\pi^4} \, a^2 \Biggl\{ \ln(a) \partial^2 \!-\!
\Bigl(2 \ln(a) \!+\! 1\Bigr) H a \partial_0 \!-\! \Bigl(2 \ln(a) \!+\!
\frac32\Bigr) H^2 a^2 \Biggr\} \delta^4(x \!-\! x') \nonumber \\
& & - \frac{\l^2 H^2}{2^7 \pi^4} \Bigl( \frac{2 \pi}{H \mu}
\Bigr)^\e \Bigg\{ \frac{(1 \!-\! \frac{\epsilon}4) (1 \!-\!
\frac{\e}2) \G^2(1 \!-\! \frac{\e}2)}{\e^2 (1 \!-\! \frac32 \e)}
\Bigl(\frac{H}{2\mu}\Bigr)^{\epsilon} \!-\! \frac{\G(1 \!-\!
\e)}{\e} \pi \cot\left(\frac{\pi \epsilon}{2}\right)
\Biggr\} a^4\delta^D(x \!-\! x') \nonumber \\
& & + \frac{\lambda^2 H^2}{2^7 \pi^4} \Bigl(\frac{2 \pi}{H
\mu}\Bigr)^{ \epsilon} \, \frac{\ln(a)}{\epsilon} \, a^4
\delta^D(x \!-\! x') \!-\! \frac{ \lambda^2 H^2}{2^7 \pi^4}
\Biggl\{ \frac12 \ln^2(a) \!+\! \left[\frac34 \!-\!
\gamma+\ln\left(\frac{H}{2\mu}\right)\right] \ln(a)\Biggr\} a^4 \delta^4(x \!-\! x') \nonumber \\
& & + \frac{i \l^2 a a'}{2^9 \pi^6} \Bigg\{\frac1{24} \dd^4
\Biggl[ \frac{\ln\left(\mu^2 \D x^2_{\scriptscriptstyle
--}\right)}{\D x^2_{ \scriptscriptstyle --}} \Biggr] \!-\! H^2 a
a' \dd^2 \Bigg[ \ln\Bigl( \frac{He^{\frac34}}{2 \mu} \Bigr)
\frac{\ln\left(\mu^2 \D x^2_{ \scriptscriptstyle --}\right)}{\D
x^2_{\scriptscriptstyle --}} \!+\! \frac{\ln^2\left(\mu^2 \D
x^2_{\scriptscriptstyle --} \right)}{4 \D
x^2_{\scriptscriptstyle --}} \Bigg] \nonumber \\
& & \hspace{3.5cm} - H^4 (a a')^2 \frac{\ln^2\left(\frac{\sqrt{e}}4 H^2
\D x^2_{\scriptscriptstyle --}\right)}{\D x^2_{\scriptscriptstyle --}} \!+\!
\frac{H^6}6 (a a')^3 \ln^3\Bigl(\frac{\sqrt{e}}4 H^2 \D x^2_{\scriptscriptstyle
--}\Bigr) \Bigg\}\; . \label{Mss--}
\ee
The mixed polarities lack the local terms and differ by a sign,
\be
& & M^2_{\mbox{{\tiny ss$+-$}}} = \frac{i \l^2}{6}
(aa')^{4-\e}[i\D_{+-}(x;x')]^3=-i\frac{\l^2 a a'}{2^9
\pi^6}\Bigg\{\frac{1}{24}\dd^4\left[\frac{\ln\left(\mu^2\D
x^2_{+-}\right)}{\D x^2_{+-}}\right]\nonumber\\&&-H^2a a'\dd^2
\left[\ln\left(\frac{He^{\frac{3}{4}}}{2\mu}\right)\frac{\ln\left(\mu^2\D
x^2_{+-}\right)}{\D x^2_{+-}}+\frac{1}{4}\frac{\ln^2\left(\mu^2\D
x^2_{+-}\right)}{\D x^2_{+-}}\right]-H^4(a
a')^2\frac{\ln^2\left(\frac{\sqrt{e}}{4}H^2\D x^2_{+-}\right)}{\D
x^2_{+-}}\nonumber\\&&\hspace{8.5cm}+\frac{H^6}{6}(a
a')^3\ln^3\left(\frac{\sqrt{e}}{4}H^2\D x^2_{+-}\right)\Bigg\} \; , \\
& & M^2_{\mbox{{\tiny ss$-+$}}}= \frac{i \l^2}{6}
(aa')^{4-\e}[i\D_{-+}(x;x')]^3=-i\frac{\l^2 a a'}{2^9
\pi^6}\Bigg\{\frac{1}{24}\dd^4\left[\frac{\ln\left(\mu^2\D
x^2_{-+}\right)}{\D x^2_{-+}}\right]\nonumber\\&&-H^2a a'\dd^2
\left[\ln\left(\frac{He^{\frac{3}{4}}}{2\mu}\right)\frac{\ln\left(\mu^2\D
x^2_{-+}\right)}{\D x^2_{-+}}+\frac{1}{4}\frac{\ln^2\left(\mu^2\D
x^2_{-+}\right)}{\D x^2_{-+}}\right]-H^4(a
a')^2\frac{\ln^2\left(\frac{\sqrt{e}}{4}H^2\D x^2_{-+}\right)}{\D
x^2_{-+}}\nonumber\\&&\hspace{8.5cm}+\frac{H^6}{6}(a
a')^3\ln^3\left(\frac{\sqrt{e}}{4}H^2\D x^2_{-+}\right)\Bigg\} \; .
\ee

\section{The Snowman Diagram}

\begin{figure}[htbp]
\centerline{\epsfig{file=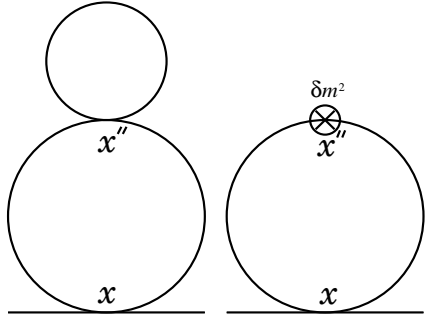,
                               width=2.2in,height=1.65in }}
\caption{Generic Snowman Diagram with mass counterterm. \label{fig4} }
\end{figure}

The primitive two loop graph is known as the snowman (sm) diagram. Unlike
the sunset diagram, it contributes only to $M^2_{\scriptscriptstyle ++}$ and
$M^2_{\scriptscriptstyle --}$. Its generic topology is depicted in
Fig.~\ref{fig4}. The right hand diagram depicts the 1-loop mass counterterm
which naturally combines with it.

Because both external points of this diagram must have the same
polarity we denote the contributions with a single subscript sign,
$M^2_{\mbox{{\tiny sm+}}}$ and $M^2_{\mbox{{\tiny sm-}}}$. The
internal vertices are summed over both polarities, as always in
the Schwinger-Keldysh formalism. A simple application of the
Feynman rules gives, \be
\!\!\!\!\!\!\!\!\!\!\!\!\!\!\!\!\!\!\!\!\!\!\!\! & &
\!\!\!\!\!\!\!\!\!\!\!\!\!\!\!\!\! \lefteqn{-iM^2_{\mbox{{\tiny
sm+}}} \!\! = \!\!\frac{-i\l a^D}{2} \d^D(x-x') \!\! \int d^Dx''
a''^D\Biggl\{
[i\D_{++}(x;x'')]^2-[i\D_{+-}(x;x'')]^2\Biggr\}}\nonumber\\&&\hspace{8.5cm}
\times\Bigg\{ \frac{(-i\l)}{2}i\D(x'';x'')-i\d m^2\Bigg\} \; .\ee
Since both diagrams of Fig.~\ref{fig4} have the same lower loop,
they possess a common factor in the first curly bracket. The first
term within the final curly bracket derives from the left hand
diagram, while the second term comes from the right hand diagram.

The coincident propagator and the mass counter term are calculated in
(\ref{coin}) and (\ref{massct}) respectively. Once they are inserted into
the above equation, one finds
\be
-iM^2_{\mbox{{\tiny sm+}}} \!\! = \!\! \frac{-\l^2 H^{2-\e} a^{4-\e}}{2^{5-\e}
\pi^{2-\frac{\e}{2}}} \frac{\G(3\!-\!\e)}{\G(2\!-\!\frac{\e}{2})}
\d^D(x\!-\!x') \!\! \int \!\! d^4x'' a''^{4-\e}\!\ln(a'')\Bigg\{\!
[i\D_{++}(x;x'')]^2\!-\![i\D_{+-}(x;x'')]^2\!\Bigg\}\; .
\nonumber \label{2lnew} \!\!\!\!\!\!\!\!\!\!\! \\
\ee
The squares of the various propagators are computed in Appendix B and the
results we need are (\ref{D2++}) and (\ref{D2--})
\be
\!\!\!\!\!\!\!\!\!\!\!\!\!\!\!\!-iM^2_{\mbox{{\tiny sm+}}} \!\! & = &
\!\! \frac{i \l^2 H^2 a^4}{2^7\pi^4} \Bigl(\frac{2 \pi}{H \mu} \Bigr)^\e
\frac{\G(1 \!-\! \e)}{\e} \ln(a) \d^D(x \!-\! x') \nonumber \\
& & \!\! + \frac{\l^2 H^2a^2}{2^{10}\pi^6}\d^4(x \!-\! x')(-\dd^2_0)\int
d^4x'' a''^2\ln(a'')\left[\frac{\ln(\mu^2\D x^2_{\scriptscriptstyle ++})}{\D
x^2_{\scriptscriptstyle ++}}-\frac{\ln(\mu^2\D x^2_{\scriptscriptstyle +-})}{\D
x^2_{\scriptscriptstyle +-}}\right]\nonumber\\ &&\!\!+\frac{\l^2 H^4
a^3}{2^8\pi^6}\d^4(x-x')\int d^4x''
a''^3\ln(a'')\Bigg[\frac{\ln(\frac{H^2\sqrt{e}}{4}\D x^2_{\scriptscriptstyle
++})}{\D x^2_{\scriptscriptstyle ++}} - \frac{\ln(\frac{H^2\sqrt{e}}{4}\D
x^2_{\scriptscriptstyle +-})}{\D x^2_{\scriptscriptstyle \scriptscriptstyle
+-}}\Bigg] \nonumber \\
& & \!\!-\frac{\l^2 H^6
a^4}{2^{10}\pi^6}\d^4(x-x')\!\!\int\!\! d^4x''
a''^4\ln(a'')\Bigg[\!\ln^2\!\!\left(\!\frac{H^2\sqrt{e}}{4}\D
x^2_{\scriptscriptstyle ++}\right)\!-\!\ln^2\!\!\left(\!\frac{H^2\sqrt{e}}{4}\D
x^2_{\scriptscriptstyle +-}\right)\!\!\Bigg]\; , \label{smints} \\
& \longrightarrow & \!\! \frac{i \l^2 H^2}{2^7\pi^4} \Bigl(\frac{2 \pi}{H \mu}
\Bigr)^\e \Biggl\{\frac{\ln(a)}{\e} \!+\! \gamma \ln(a) \Biggr\} \, a^4
\d^D(x \!-\! x') \nonumber \\
& & \hspace{2cm} +
\frac{\lambda^2 H^2 a^2}{2^{10} \pi^6} \delta^4(x \!-\! x') \Bigl\{ I_1 + 4
H^2 a I_2 - H^4 a^2 I_3\Bigr\} \; .
\ee
The integrals $I_1$, $I_2$ and $I_3$ are evaluated in Appendix D and the
results are expressions (\ref{I1}), (\ref{I2}), and (\ref{I3}). Substituting
these and adding everything up gives,
\be
M^2_{\mbox{{\tiny sm+}}} \!\! & = & \!\! -\frac{\l^2 H^2}{2^7\pi^4}
\Bigl(\frac{2 \pi}{H \mu} \Bigr)^\e \, \frac{\ln(a)}{\e} \, a^4
\delta^D(x \!-\! x') + \frac{\l^2 H^2}{2^7 \pi^4} \, a^4 \d^4(x \!-\! x')
\Bigg\{-\frac49 \ln^3(a) \nonumber \\
& & \!\! + \frac{13}{18} \ln^2(a) \!+\! \Bigl[\ln\Bigl(\frac{H}{2 \mu}\Bigr)
\!+\! \frac83 \!-\! \gamma \!-\! \frac29 \pi^2 \Bigr] \ln(a) \!-\!
\frac{238}{81} \!+\!  \frac{13}{54} \pi^2 \!+\! \frac43 \zeta(3) \!+\!
\frac{a^{-3}}{81} \nonumber \\
& & \!\! - \sum_{n=1}^\infty \frac{n+5}{(n+1)^3} a^{-(n+1)} \!+\! 4
\sum_{n=1}^\infty\frac{a^{-(n+2)}}{(n+2)^3} \!+\! 4 \sum_{n=1}^\infty
\frac{a^{-(n+3)}}{n(n+3)^3} \Bigg\} \label{SMa} \; . \label{Msm++}
\ee
Note that the overlapping divergence from the first term is the same as the
one from the sunset diagram. The ${\scriptscriptstyle --}$ case is minus
(\ref{Msm++}).

\section{The Counter Diagrams}

\begin{figure}[htbp]
\centerline{\epsfig{file=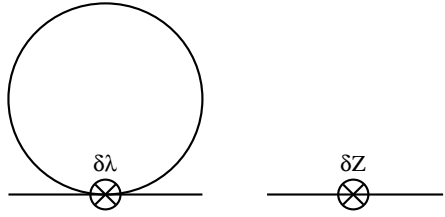,
                               width=2.3in,height=1.1in }}
\caption{Generic Counter Diagrams. The one on the left is the same as
Fig.~\ref{fig1} but with $\delta \lambda$ in place of $\lambda$.
The one on the right is the field strength renormalization. \label{fig5} }
\end{figure}

We have already encountered the mass counterterm in Section III. At the
order we are working there is also a one loop diagram with $\lambda$
replaced by $\delta \lambda$, and there is a field strength renormalization.
The generic diagram topologies are depicted in Fig.~\ref{fig5}. Note that
neither diagram contributes to the mixed polarities.

The contribution of the one loop counter vertex diagram to the
${\scriptscriptstyle ++}$ polarity is,
\be
-iM^2_{\mbox{{\tiny cv+}}} & = & -i \frac{\d\l}2 a^D i \D(x';x') \d^D(x-x')
\; . \label{cv}
\ee
Of course we can read off the answer from Section III,
\be
M^2_{\mbox{{\tiny cv+}}} & = & \frac{3 H^2 \l^2 a^{4-\e}}{2^8 \pi^4}
\Bigl(\frac{2\pi}{\mu H}\Bigr)^\e \, \frac{\G(1 \!-\! \e)}{\e} \Bigg\{ 2 \ln(a)
+ \pi \cot\Bigl(\frac{\pi\e}{2}\Bigr) \Bigg\} \d^D(x \!-\! x') \; , \\
& \longrightarrow & \frac{H^2 \lambda^2}{2^7 \pi^4} \Bigl(\frac{2 \pi}{\mu H}
\Bigr)^{\epsilon} \left\{\frac32 \frac{\Gamma(1 \!-\! \epsilon)}{\epsilon}
\pi \cot\Bigl(\frac{\pi \epsilon}2 \Bigr) - \frac32 \ln^2(a) \right\} a^4
\d^D(x \!-\! x') \; . \label{Mcv++}
\ee
The ${\scriptscriptstyle --}$ case is minus this.

The contribution of field strength renormalization to the
${\scriptscriptstyle ++}$ polarity is,
\be
M^2_{\mbox{{\tiny $\delta Z$+} }} = - \delta Z \eta^{\mu\nu} \dd_\mu \Bigl(
a^{D-2} \dd_\nu \d^D(x \!-\! x') \Bigr) = - \d Z a^{2-\e} \Bigl( \dd^2 \!-\!
(2 \!-\! \e) H a \dd_0 \Bigr) \d^D(x \!-\! x') \; .
\ee
The only primitive graph with a divergence proportional to $a^2 \Bigl(
\partial^2 \!-\! 2 H a \partial_0\Bigr) \delta^D(x-x')$ is the sunset diagram.
Comparison with our result (\ref{Mss++}) for it implies, \be \d Z
= - \frac{\l^2}{2^{10} 3 \pi^4} \!
\Bigl(\frac{\pi}{\mu^2}\Bigr)^\e \, \frac{\G^2(1 \!-\!
\frac{\e}2)}{(1 \!-\! \frac34\e) (1 \!-\! \frac32 \e) (1 \!-\! \e)
\e} \; . \ee Although some finite, local terms result from
expanding $a^{-\epsilon} = 1 - \epsilon \ln(a) + O(\epsilon^2)$
there are no overlapping divergences from the field strength
renormalization at two loop order, \be M^2_{\mbox{{\tiny $\delta
Z$+} }} & = & \frac{\l^2}{2^{10} 3 \pi^4} \! \Bigl(
\frac{\pi}{\mu^2} \Bigr)^\e \, \frac{\G^2(1 \!-\!
\frac{\e}{2})}{(1 \!-\! \frac34\e) (1 \!-\! \frac32 \e) (1 \!-\!
\e) \e} \, a^2 \Bigl( \dd^2
\!-\!  2 H a \dd_0 \Bigr) \d^D(x \!-\! x') \nonumber \\
& & \hspace{2cm} + \frac{\l^2}{2^{10} 3 \pi^4} \, a^2 \left[ -\ln(a) \partial^2
\!+\! \Bigl(2 \ln(a) \!+\! 1\Bigr) H a \partial_0 \right] \delta^4(x \!-\!x')
\; . \label{Mz++}
\ee

After field strength renormalization, all remaining divergences
reside on $a^4 \d^D(x \!-\!x')$. Such contributions comes from the
sunset diagram (\ref{Mss++}) and the counter vertex (\ref{Mcv++}).
Comparison with these results suggests that the simplest choice
for the two loop mass counterterm is, \be \d m^2
\Bigl\vert_{\lambda^2} & = & -\frac{\l^2H^2}{2^7 \pi^4}
\Bigl(\frac{2 \pi}{H \mu}\Bigr)^{\epsilon} \, \Bigg\{
\Bigl(\frac{H}{2 \mu}\Bigr)^{\epsilon} \frac{\Gamma^2(1 \!-\!
\frac{\epsilon}2)}{12 (1 \!-\! \frac32 \epsilon) (1 \!-\! \frac34
\e) (1 \!-\!  \e) \e} \!+\! \Bigl(\frac{H}{2 \mu} \Bigr)^\e \,
\frac{(1 \!-\! \frac{\epsilon}4) (1 \!-\!\frac{\epsilon}{2})
\Gamma^2(1 \!-\!
\frac{\epsilon}2)}{(1 \!-\! \frac32 \epsilon) \epsilon^2} \nonumber \\
& & \hspace{2cm} + \frac{\G(1 \!-\!\e)}{2 \e}
\pi\cot{\Bigl(\frac{\pi \e}2\Bigr)} +\frac1{16} - \frac{238}{81} +
\frac{13}{54} \pi^2 + \frac43 \zeta(3) \Bigg\} \; . \ee Of course
the mass counterterm is $M^2_{\scriptscriptstyle \delta m +} =
\delta m^2 a^{4 - \epsilon} \delta^D(x \!-\! x')$. Using the
expansion $a^{-\epsilon} = 1 \!-\! \epsilon \ln(a) \!+\! \frac12
\epsilon^2 \ln^2(a) \!+\!  O(\epsilon^3)$ we see that the two loop
contribution is,
\begin{eqnarray}
\lefteqn{M^2_{\mbox{{\tiny $\delta m$+} }} \!\! \Biggl\vert_{\lambda^2}
\!\!\!\!\!\ \longrightarrow -
\frac{\lambda^2 H^2}{2^7 \pi^4} \Bigl(\frac{2 \pi}{H \mu}\Bigr)^{\epsilon}
\Bigg\{\Bigl(\frac{H}{2 \mu} \Bigr)^\e \, \frac{\Gamma^2(1 \!-\!
\frac{\epsilon}2)}{12 (1 \!-\! \frac32 \epsilon) (1 \!-\! \frac34 \e) (1 \!-\!
\e) \e} \!+\! \Bigl(\frac{H}{2 \mu} \Bigr)^\e \, \frac{(1 \!-\!
\frac{\epsilon}4) (1 \!-\! \frac{\epsilon}2) \Gamma^2(1 \!-\!
\frac{\epsilon}2)}{(1 \!-\! \frac32 \epsilon) \epsilon^2}} \nonumber \\
& & + \frac{\Gamma(1 \!-\! \e)}{2 \epsilon} \pi
\cot{\Bigl(\frac{\pi \epsilon} 2\Bigr)} \Biggr\} a^4 \delta^D(x
\!-\! x') + \frac{\lambda^2 H^2}{2^6 \pi^4} \Bigl(\frac{2 \pi}{H
\mu}\Bigr)^{
\epsilon} \, \frac{\ln(a)}{\epsilon} \, a^4 \delta^D(x \!-\! x') \nonumber \\
& & - \frac{\lambda^2 H^2}{2^7 \pi^4} \Biggl\{ \ln^2(a) \!-\!
\Biggl[\frac56 \!+\! 2 \gamma \!+\! \ln\Bigl(\frac{H}{2
\mu}\Bigr)\Biggr] \ln(a) \!+\! \frac1{16} \!-\! \frac{238}{81}
\!+\! \frac{13}{54} \pi^2 \!+\! \frac43 \zeta(3) \Bigg\} a^4
\delta^4(x \!-\! x') \; . \label{Mm++}
\end{eqnarray}
Note that the nonlocal divergence precisely cancels the nonlocal divergences
from the sunset diagram (\ref{Mss++}) and the snowman diagram (\ref{Msm++}).
As usual, the ${\scriptscriptstyle --}$ polarity is minus (\ref{Mm++}).

\section{The Fully Renormalized Result}

To get the full two loop result for the ${\scriptscriptstyle ++}$ case we
add the contributions from the sunset diagram (\ref{Mss++}), the snowman
diagram (\ref{Msm++}), the counter vertex diagram (\ref{Mcv++}), the field
strength renormalization (\ref{Mz++}) and the two loop mass renormalization
(\ref{Mm++}),
\be
& & \!\!\!\!\!\!M^2_{\mbox{{\tiny 2++}}} \!=\! M^2_{\mbox{{\tiny ss++}}}
+ M^2_{\mbox{{\tiny sm+}}} + M^2_{\mbox{{\tiny cv+}}} +
M^2_{\mbox{{\tiny $\delta Z$+}}} + M^2_{\mbox{{\tiny $\delta m$+}}}\nonumber \\
& & \!\!\!\!\!\! = \! \frac{i \l^2}{2^9 \pi^6} \Bigg\{\frac{a a'}{24} \dd^4
\Biggl[\frac{\ln\left(\mu^2 \D x^2_{\scriptscriptstyle ++}\right)}{\D
x^2_{\scriptscriptstyle ++}} \Biggr] \!-\! H^2 (a a')^2 \dd^2 \Biggl[
\ln\Bigl(\frac{H e^{\frac34}}{2 \mu}\!\Bigr) \frac{\ln\left(\mu^2 \D x^2_{
\scriptscriptstyle ++}\right)}{\D x^2_{\scriptscriptstyle ++}} \!+\!
\frac{\ln^2\left(\mu^2 \D x^2_{\scriptscriptstyle ++}\right)}{4 \D x^2_{
\scriptscriptstyle ++}}\Biggr] \nonumber \\
& & - H^4 (a a')^3 \frac{ \ln^2\left(\frac{\sqrt{e}}4 H^2 \D
x^2_{\scriptscriptstyle ++}\right)}{\D x^2_{\scriptscriptstyle ++}} \!+\!
\frac{H^6}6 (a a')^4 \! \ln^3\Bigl(\frac{\sqrt{e}}4 H^2 \D x^2_{
\scriptscriptstyle ++} \Bigr) \Bigg\} \nonumber \\
& & + \frac{\l^2}{2^9 3\, \pi^4} \, a^2 \Biggl\{-\ln(a) \dd^2 \!+\!
\Bigl(2 \ln(a) \!+\! 1\Bigr) H a \partial_0 \Biggr\} \delta^4(x \!-\! x')
\nonumber \\
& & + \frac{\lambda^2 H^2}{2^7 \pi^4} \Biggl\{-\frac49 \ln^3(a)
\!-\! \frac{23}{18} \ln^2(a) \!+\! \Biggl[\frac{13}3 \!+\! 3
\ln\Bigl(\frac{H}{2 \mu} \Bigr) \!-\! \frac29 \pi^2 \Biggr] \ln(a)
\Biggl\} a^4 \delta^4(x \!-\! x')
\nonumber \\
& & + \frac{\lambda^2 H^2}{2^7 \pi^4} \Biggl\{ \frac{a^{-3}}{81} \!-\!
\sum_{n=1}^\infty \frac{n+5}{(n+1)^3}a^{-(n+1)} \!+\! 4 \sum_{n=1}^\infty
\frac{a^{-(n+2)}}{(n+2)^3} \!+\! 4 \sum_{n=1}^\infty\frac{a^{-(n+3)}}{n(n+3)^3}
\Biggr\} a^4 \delta^4(x \!-\! x') \; . \label{M2++}
\ee
As with the renormalized stress-energy tensor \cite{OW2}, we conjecture
that the exponentially falling terms on the last line could be absorbed
into a perturbative modification of the initial state. The ${
\scriptscriptstyle --}$ polarity is minus the complex conjugate,
\be
& & \!\!\!\!\!\! M^2_{\mbox{{\tiny 2$--$}}} \!=\! -\Bigl(M^2_{\mbox{{\tiny
2++}}} \Bigr)^* \nonumber \\
& & \!\!\!\!\!\! = \! \frac{i \l^2}{2^9 \pi^6} \Bigg\{\frac{a a'}{24} \dd^4
\Biggl[\frac{\ln\left(\mu^2 \D x^2_{\scriptscriptstyle --}\right)}{\D
x^2_{\scriptscriptstyle --}} \Biggr] \!-\! H^2 (a a')^2 \dd^2 \Biggl[
\ln\Bigl(\frac{H e^{\frac34}}{2 \mu}\!\Bigr) \frac{\ln\left(\mu^2 \D x^2_{
\scriptscriptstyle --}\right)}{\D x^2_{\scriptscriptstyle --}} \!+\!
\frac{\ln^2\left(\mu^2 \D x^2_{\scriptscriptstyle --}\right)}{4 \D x^2_{
\scriptscriptstyle --}}\Biggr] \nonumber \\
& & - H^4 (a a')^3 \frac{ \ln^2\left(\frac{\sqrt{e}}4 H^2 \D
x^2_{\scriptscriptstyle --}\right)}{\D x^2_{\scriptscriptstyle --}} \!+\!
\frac{H^6}6 (a a')^4 \! \ln^3\Bigl(\frac{\sqrt{e}}4 H^2 \D x^2_{
\scriptscriptstyle --} \Bigr) \Bigg\} \nonumber \\
& & - \frac{\l^2}{2^9 3\, \pi^4} \, a^2 \Biggl\{-\ln(a) \dd^2 \!+\!
\Bigl(2 \ln(a) \!+\! 1\Bigr) H a \partial_0 \Biggr\} \delta^4(x \!-\! x')
\nonumber \\
& & - \frac{\lambda^2 H^2}{2^7 \pi^4} \Biggl\{-\frac49 \ln^3(a)
\!-\! \frac{23}{18} \ln^2(a) \!+\! \Biggl[\frac{13}3 \!+\! 3
\ln\Bigl(\frac{H}{2 \mu} \Bigr) \!-\! \frac29 \pi^2 \Biggr] \ln(a)
\Biggl\} a^4 \delta^4(x \!-\! x')
\nonumber \\
& & - \frac{\lambda^2 H^2}{2^7 \pi^4} \Biggl\{ \frac{a^{-3}}{81}
\!-\! \sum_{n=1}^\infty \frac{n+5}{(n+1)^3}a^{-(n+1)} \!+\! 4
\sum_{n=1}^\infty \frac{a^{-(n+2)}}{(n+2)^3} \!+\! 4
\sum_{n=1}^\infty\frac{a^{-(n+3)}}{n(n+3)^3} \Biggr\} a^4
\delta^4(x \!-\! x') \; . \label{m--} \ee As usual, the
${\scriptscriptstyle +-}$ and ${\scriptscriptstyle -+}$ polarities
have the opposite sign and lack the local terms, \be
\!\!\!\!\!\!M^2_{\mbox{{\tiny2$+-$}}}\!\!&=&\!\! -\frac{i
\l^2}{2^9 \pi^6} \Bigg\{\frac{a a'}{24} \dd^4
\Biggl[\frac{\ln\left(\mu^2 \D x^2_{\scriptscriptstyle
+-}\right)}{\D x^2_{\scriptscriptstyle +-}} \Biggr] \!-\! H^2 (a
a')^2 \dd^2 \Biggl[ \ln\Bigl(\frac{H e^{\frac34}}{2 \mu}\!\Bigr)
\frac{\ln\left(\mu^2 \D x^2_{ \scriptscriptstyle +-}\right)}{\D
x^2_{\scriptscriptstyle +-}} \!+\! \frac{\ln^2\left(\mu^2 \D
x^2_{\scriptscriptstyle +-}\right)}{4 \D x^2_{
\scriptscriptstyle + -}}\Biggr] \nonumber \\
\!\!&&\!\!- H^4 (a a')^3 \frac{ \ln^2\left(\frac{\sqrt{e}}4 H^2 \D
x^2_{\scriptscriptstyle +-}\right)}{\D x^2_{\scriptscriptstyle
+-}} \!+\! \frac{H^6}6 (a a')^4 \! \ln^3\Bigl(\frac{\sqrt{e}}4 H^2
\D x^2_{ \scriptscriptstyle +-} \Bigr) \Bigg\}\\
\!\!\!\!\!\!M^2_{\mbox{{\tiny 2$-+$}}}\!\!&=&\!\! -\frac{i
\l^2}{2^9 \pi^6} \Bigg\{\frac{a a'}{24} \dd^4
\Biggl[\frac{\ln\left(\mu^2 \D x^2_{\scriptscriptstyle
-+}\right)}{\D x^2_{\scriptscriptstyle -+}} \Biggr] \!-\! H^2 (a
a')^2 \dd^2 \Biggl[ \ln\Bigl(\frac{H e^{\frac34}}{2 \mu}\!\Bigr)
\frac{\ln\left(\mu^2 \D x^2_{ \scriptscriptstyle -+}\right)}{\D
x^2_{\scriptscriptstyle -+}} \!+\! \frac{\ln^2\left(\mu^2 \D
x^2_{\scriptscriptstyle -+}\right)}{4 \D x^2_{
\scriptscriptstyle -+}}\Biggr] \nonumber \\
\!\!&&\!\!- H^4 (a a')^3 \frac{ \ln^2\left(\frac{\sqrt{e}}4 H^2 \D
x^2_{\scriptscriptstyle -+}\right)}{\D x^2_{\scriptscriptstyle
+-}} \!+\! \frac{H^6}6 (a a')^4 \! \ln^3\Bigl(\frac{\sqrt{e}}4 H^2
\D x^2_{ \scriptscriptstyle -+} \Bigr) \Bigg\} \ee

\section{Discussion}

We have computed and fully renormalized the four Schwinger-Keldysh
self-mass-squared's at one and two loop orders for massless, minimally
coupled $\varphi^4$ (\ref{simL}) on a locally de Sitter background
(\ref{ds^2}). The next step is using our results to study the
quantum-corrected, effective field equations at linearized order
(\ref{linord}). The term ``$M^2(x;x')$'' called for in (\ref{linord}) is
the sum of $M^2_{\scriptscriptstyle ++}(x;x')$ and $M^2_{\scriptscriptstyle
+-}(x;x')$. The analysis --- which will be presented in a subsequent paper
--- is very similar to that already done for de Sitter photons using the
one loop vacuum polarization \cite{PW1} and for de Sitter fermions using
the one loop self-energy \cite{PW2}.

There are two interesting issues to investigate. The first is stability.
Secular effects in this system are driven by inflationary particle
production which forces the scalar up its $\varphi^4$ potential. One
would expect this to generate a {\it positive} mass-squared, and it is
significant that our one loop result (\ref{M1++}) is positive. The two
loop result (\ref{M2++}) is a little too complicated to interpret
without a detailed calculation of the type proposed, but we expect no
problems with stability.

The second interesting issue is an apparent paradox in the stochastic
solution of this model which was given in 1994 by Starobinski\u{\i} and
Yokoyama \cite{SY}. As we noted in the introduction, this model is
stable for two reasons. First, as inflationary particle production forces the
scalar up its potential the classical force pushes it back down. Second,
the curvature of the potential ought to induce a positive mass-squared
which should reduce the inflationary particle production. However, the
solution of Starobinski\u{\i} and Yokoyama implements only the first effect.
Although it would be possible to implement the second effect as well
within the context of Starobinski\u{\i}'s stochastic technique \cite{Staro},
including just the first effect gives perfect agreement at leading
logarithm order as high as this has been checked. The question is why
should this be so? A possible answer is that the nonzero self-mass-squared
remains always down by one factor of $\ln(a)$. Whereas the leading
logarithms of the energy density give a series in $\lambda \ln^2(a)$,
\begin{equation}
\rho \Bigl\vert_{\rm leading} = H^4 \sum_{\ell = 2}^{\infty}
\rho_{\ell} \Bigl(\lambda \ln^2(a) \Bigr)^{\ell -1} \; ,
\end{equation}
it may be that the self-mass-squared behaves as such a series times
$\lambda \ln(a)$,
\begin{equation}
M^2(x;x') \Bigl\vert_{\rm leading} = H^2 a^4 \delta^4(x \!-\! x') \times
\lambda \ln(a) \sum_{\ell = 1}^{\infty} m_{\ell} \Bigl(\lambda \ln^2(a)
\Bigr)^{\ell-1} \; .
\end{equation}

It is worth noting that, had gravity been quantized, there would have been
similar corrections from graviton exchange. While the physics of the pure
scalar model pretty much guarantees that a positive self-mass-squared must
be induced, there is no such argument for gravity. If the induced
self-mass-squared turns out to be tachyonic it might have important implications
for scalar-driven inflation. Further, there is no reason the gravitational
process should ever stop. The graviton should not experience a classical force
since it has no non-derivative potential. Nor should the graviton ever
develop a mass. This might be significant for scalar-driven models which
inflate for a very large number of e-foldings.

We close with a comment on accuracy. Two loop computations are generally
intricate enough that this becomes an issue. In this regard it is worth
noting that three distinct diagrams conspire to cancel the overlapping
divergences. This is significant because there are no overlapping divergences
at this order in massless $\varphi^4$ on flat space.

It is also a significant correspondence check that we get the
same results for $\delta \lambda$ (at one loop) and $\delta Z$ (at two
loops) as for massless $\varphi^4$ in flat space. This had to be the
case because the divergences these counterterms absorb are formally
logarithmic. Because geometric corrections make divergences more finite,
any geometric corrections to $\delta \lambda$ and $\delta Z$ must be
finite. The same is not the case for $\delta m^2$ because the divergences
it cancels are formally quadratic. Indeed, $\delta m^2$ vanishes for
massless $\varphi^4$ in flat space whereas we found logarithmically
divergent corrections proportional to $H^2$ at both one and two loop
orders.

\section{Appendix}

\subsection{Partial Integration Identities}

In almost all cases derivatives of the conformal coordinate intervals
${\Delta x}^2_{\scriptscriptstyle \pm\pm}(x;x')$ can be taken generically,
without regard to polarity and without worrying about delta functions,
\begin{equation}
\partial^2 \left( \frac1{\Delta x^{n - m \epsilon}_{\scriptscriptstyle \pm\pm}}
\right) = \frac{\Bigl( n \!-\! m \epsilon\Bigr) \Bigl(n \!-\! 2 \!-\!
(m \!-\! 1) \epsilon\Bigr)}{\Delta x^{n + 2 - m \epsilon}_{\scriptscriptstyle
\pm \pm}} \qquad \qquad \Bigl(n \neq 2 \;\; {\rm or} \;\; m \neq 1\Bigr) .
\end{equation}
The exception is $1/{\Delta x}^{2 - \epsilon}$,
\begin{equation}
\partial^2 \left(\frac1{\Delta x^{2-\e}_{++}}\right) =
\frac{i 4 \pi^{2-\frac{\e}2}}{\Gamma(1 \!-\! \frac{\e}{2})}
\, \delta^D(x \!-\! x') = -\partial^2 \left(\frac1{\Delta x^{2-\e}_{--}}
\right) \;\;\; , \;\;\; \partial^2 \left(\frac1{\Delta x^{2-\e}_{+-}}\right)
= 0 = \partial^2 \left(\frac1{\Delta x^{2-\e}_{- +}} \right) . \label{delta}
\end{equation}
These identities allow one to extract more and more derivatives of inverse
powers of $\Delta x^2_{\scriptscriptstyle \pm\pm}$ until the result is
integrable in $D=4$ dimensions. For $m \neq 1$ we have,
\begin{eqnarray}
\frac1{\Delta x^{4 - m\epsilon}_{\scriptscriptstyle \pm\pm}} & = &
\frac{\partial^2}{\Bigl(2 \!-\! m \epsilon\Bigr) \Bigl(- (m \!-\! 1)
\epsilon\Bigr)} \left(\frac1{\Delta x^{2 - m \epsilon}_{\pm\pm}} \right)
\; , \\
\frac1{\Delta x^{6 - m\epsilon}_{\scriptscriptstyle \pm\pm}} & = &
\frac{\partial^4}{\Bigl(4 \!-\! m \epsilon\Bigr) \Bigl(2 \!-\! m\epsilon\Bigr)
\Bigl(2 \!-\! (m \!-\! 1) \epsilon\Bigr) \Bigl(- (m \!-\! 1) \epsilon\Bigr)}
\left(\frac1{\Delta x^{2 - m \epsilon}_{\pm\pm}} \right) \; .
\end{eqnarray}
At this stage one adds zero in the form of the appropriate identity
(\ref{delta}) and then expands the nonlocal term in powers of $\epsilon$.
For example, the ${\scriptscriptstyle ++}$ term gives,
\begin{eqnarray}
\partial^2 \left(\frac1{\Delta x^{2 - m \epsilon}_{\scriptscriptstyle ++}}
\right) & = & \partial^2 \left(\frac1{\Delta x^{2 - m \epsilon}_{
\scriptscriptstyle ++}} - \frac{\mu^{-(m-1)\epsilon}}{\Delta x^{2 - \epsilon}_{
\scriptscriptstyle ++}} \right) + \frac{i 4 \pi^{2-\frac{\epsilon}2} \mu^{-(m
-1) \epsilon}}{\Gamma(1 \!-\!  \frac{\epsilon}{2})} \, \delta^D(x \!-\! x')
\; , \\
& = & + \frac{i 4 \pi^{2-\frac{\epsilon}2} \mu^{-(m -1) \epsilon}}{\Gamma(1
\!-\!  \frac{\epsilon}{2})} \, \delta^D(x \!-\! x') \nonumber \\
& & + \mu^{-m \epsilon} \partial^2 \left( \frac{\frac{(m \!-\! 1)}2 \epsilon
\ln\Bigl(\mu^2 \Delta x^2_{\scriptscriptstyle ++}\Bigr) \!+\! \frac{(m^2 \!-\!
1)}8 \epsilon^2 \ln^2\Bigl(\mu^2 \Delta x^2_{\scriptscriptstyle ++}\Bigr)
+ O(\epsilon^3)}{\Delta x^2_{\scriptscriptstyle ++}}\right) . \qquad
\end{eqnarray}

The cases we require are,
\begin{eqnarray}
\frac1{\Delta x^{4-2\epsilon}_{\scriptscriptstyle ++}} & = & \frac{\mu^{-
\epsilon}}{(2 \!-\! 2 \epsilon) (-\epsilon)} \, \frac{i 4 \pi^{2-
\frac{\epsilon}2}}{\Gamma(1 \!-\!  \frac{\epsilon}{2})} \, \delta^D(x \!-\! x')
\nonumber \\
& & + \frac{\mu^{-2 \epsilon} \, \partial^2}{(2 \!-\! 2 \epsilon) (-\epsilon)}
\left( \frac{\frac12 \epsilon \ln\Bigl(\mu^2 \Delta x^2_{\scriptscriptstyle ++}
\Bigr) + \frac38 \epsilon^2 \ln^2\Bigl(\mu^2 \Delta x^2_{\scriptscriptstyle ++}
\Bigr) + O(\epsilon^3)}{\Delta x^2_{\scriptscriptstyle ++}} \right) \; ,
\label{++42} \\
\frac1{\Delta x^{4-3\epsilon}_{\scriptscriptstyle ++}} & = & \frac{\mu^{-2
\epsilon}}{(2 \!-\! 3 \epsilon) (-2\epsilon)} \, \frac{i 4 \pi^{2-
\frac{\epsilon}2}}{\Gamma(1 \!-\!  \frac{\epsilon}{2})} \, \delta^D(x \!-\! x')
\nonumber \\
& & + \frac{\mu^{-3 \epsilon} \, \partial^2}{(2 \!-\! 3 \epsilon) (-2\epsilon)}
\left( \frac{\epsilon \ln\Bigl(\mu^2 \Delta x^2_{\scriptscriptstyle ++}\Bigr)
+ \epsilon^2 \ln^2\Bigl(\mu^2 \Delta x^2_{\scriptscriptstyle ++} \Bigr) +
O(\epsilon^3)}{\Delta x^2_{\scriptscriptstyle ++}} \right) \; , \label{++43} \\
\frac1{\Delta x^{6-3\epsilon}_{\scriptscriptstyle ++}} & = & \frac{\mu^{-2
\epsilon}}{(4 \!-\! 3 \epsilon) (2 \!-\! 3 \epsilon) (2 \!-\! 2 \epsilon)
(-2\epsilon)} \, \frac{i 4 \pi^{2- \frac{\epsilon}2}}{\Gamma(1 \!-\!
\frac{\epsilon}{2})} \, \partial^2 \delta^D(x \!-\! x')
\nonumber \\
& & + \frac{\mu^{-3 \epsilon} \, \partial^4}{(4 \!-\! 3 \epsilon) (2 \!-\! 3
\epsilon) (2 \!-\! 2 \epsilon) (-2\epsilon)} \left( \frac{\epsilon \ln\Bigl(
\mu^2 \Delta x^2_{\scriptscriptstyle ++}\Bigr) + O(\epsilon^2)}{\Delta x^2_{
\scriptscriptstyle ++}} \right) \; . \label{++63}
\end{eqnarray}
The ${\scriptscriptstyle --}$ variations differ only in the sign of the
delta function term,
\begin{eqnarray}
\frac1{\Delta x^{4-2\epsilon}_{\scriptscriptstyle --}} & = & -\frac{\mu^{-
\epsilon}}{(2 \!-\! 2 \epsilon) (-\epsilon)} \, \frac{i 4 \pi^{2-
\frac{\epsilon}2}}{\Gamma(1 \!-\!  \frac{\epsilon}{2})} \, \delta^D(x \!-\! x')
\nonumber \\
& & + \frac{\mu^{-2 \epsilon} \, \partial^2}{(2 \!-\! 2 \epsilon) (-\epsilon)}
\left( \frac{\frac12 \epsilon \ln\Bigl(\mu^2 \Delta x^2_{\scriptscriptstyle --}
\Bigr) + \frac38 \epsilon^2 \ln^2\Bigl(\mu^2 \Delta x^2_{\scriptscriptstyle --}
\Bigr) + O(\epsilon^3)}{\Delta x^2_{\scriptscriptstyle --}} \right) \; ,
\label{--42} \\
\frac1{\Delta x^{4-3\epsilon}_{\scriptscriptstyle --}} & = & -\frac{\mu^{-2
\epsilon}}{(2 \!-\! 3 \epsilon) (-2\epsilon)} \, \frac{i 4 \pi^{2-
\frac{\epsilon}2}}{\Gamma(1 \!-\!  \frac{\epsilon}{2})} \, \delta^D(x \!-\! x')
\nonumber \\
& & + \frac{\mu^{-3 \epsilon} \, \partial^2}{(2 \!-\! 3 \epsilon) (-2\epsilon)}
\left( \frac{\epsilon \ln\Bigl(\mu^2 \Delta x^2_{\scriptscriptstyle --}\Bigr)
+ \epsilon^2 \ln^2\Bigl(\mu^2 \Delta x^2_{\scriptscriptstyle --} \Bigr) +
O(\epsilon^3)}{\Delta x^2_{\scriptscriptstyle --}} \right) \; , \label{--43} \\
\frac1{\Delta x^{6-3\epsilon}_{\scriptscriptstyle --}} & = & -\frac{\mu^{-2
\epsilon}}{(4 \!-\! 3 \epsilon) (2 \!-\! 3 \epsilon) (2 \!-\! 2 \epsilon)
(-2\epsilon)} \, \frac{i 4 \pi^{2- \frac{\epsilon}2}}{\Gamma(1 \!-\!
\frac{\epsilon}{2})} \, \partial^2 \delta^D(x \!-\! x')
\nonumber \\
& & + \frac{\mu^{-3 \epsilon} \, \partial^4}{(4 \!-\! 3 \epsilon) (2 \!-\! 3
\epsilon) (2 \!-\! 2 \epsilon) (-2\epsilon)} \left( \frac{\epsilon \ln\Bigl(
\mu^2 \Delta x^2_{\scriptscriptstyle --}\Bigr) + O(\epsilon^2)}{\Delta x^2_{
\scriptscriptstyle --}} \right) \; . \label{--63}
\end{eqnarray}
And the mixed terms have no delta functions,
\begin{eqnarray}
\frac1{\Delta x^{4-2\epsilon}_{\scriptscriptstyle +-}} & = &
\frac{\mu^{-2 \epsilon} \, \partial^2}{(2 \!-\! 2 \epsilon) (-\epsilon)}
\left( \frac{\frac12 \epsilon \ln\Bigl(\mu^2 \Delta x^2_{\scriptscriptstyle +-}
\Bigr) + \frac38 \epsilon^2 \ln^2\Bigl(\mu^2 \Delta x^2_{\scriptscriptstyle +-}
\Bigr) + O(\epsilon^3)}{\Delta x^2_{\scriptscriptstyle +-}} \right) \; ,
\label{+-42} \\
\frac1{\Delta x^{4-3\epsilon}_{\scriptscriptstyle +-}} & = &
\frac{\mu^{-3 \epsilon} \, \partial^2}{(2 \!-\! 3 \epsilon) (-2\epsilon)}
\left( \frac{\epsilon \ln\Bigl(\mu^2 \Delta x^2_{\scriptscriptstyle +-}\Bigr)
+ \epsilon^2 \ln^2\Bigl(\mu^2 \Delta x^2_{\scriptscriptstyle +-} \Bigr) +
O(\epsilon^3)}{\Delta x^2_{\scriptscriptstyle +-}} \right) \; , \label{+-43} \\
\frac1{\Delta x^{6-3\epsilon}_{\scriptscriptstyle +-}} & = &
\frac{\mu^{-3 \epsilon} \, \partial^4}{(4 \!-\! 3 \epsilon) (2 \!-\! 3
\epsilon) (2 \!-\! 2 \epsilon) (-2\epsilon)} \left( \frac{\epsilon \ln\Bigl(
\mu^2 \Delta x^2_{\scriptscriptstyle +-}\Bigr) + O(\epsilon^2)}{\Delta x^2_{
\scriptscriptstyle +-}} \right) \; . \label{+-63}
\end{eqnarray}
Replacing ${\scriptscriptstyle +-}$ with ${\scriptscriptstyle -+}$ gives the
final case.

\subsection{The Square of a Propagator}
\label{sub:D^2}
When the square of a propagator multiplies regular
expressions we first decompose it according to (\ref{split}) and
then take $\epsilon$ to zero in all but the $A^2$ term,
\begin{eqnarray}
\Bigl(i \Delta(x;x') \Bigr)^2 & = & \Bigl( A(x;x')\Bigr)^2 + 2 A(x;x')
\Bigl( B(x;x') \!+\! C(x;x') \Bigr) + \Bigl(B(x;x') \!+\! C(x;x')\Bigr)^2
, \qquad \\
& \longrightarrow & \frac{\Gamma^2(1 \!-\! \frac{\epsilon}2)}{2^4 \pi^{4 -
\epsilon}} \frac{(a a')^{-2 + \epsilon}}{{\Delta x}^{4 - 2\epsilon}} -
\frac{H^2}{2^4 \pi^4} \frac{\ln\Bigl( \frac{\sqrt{e}}4 H^2 {\Delta x}^2
\Bigr)}{a a' {\Delta x}^2} + \frac{H^4}{2^6 \pi^4} \ln^2\Bigl( \frac{\sqrt{e}}4
H^2 {\Delta x}^2 \Bigr) .
\end{eqnarray}
We then employ the appropriate identity --- ({\ref{++42}),
(\ref{--42}) or (\ref{+-42}) --- and take $\epsilon = 0$ in the
nonlocal part. The squares of the ${\scriptscriptstyle +-}$ and
${\scriptscriptstyle +-}$ propagators are finite,
\begin{eqnarray}
\Bigl(i \Delta_{\scriptscriptstyle +-}\Bigr)^2 \!\!\!\! &
\rightarrow & \!\! - \frac{(a a')^{-2}}{2^6 \pi^4} \partial^2 \!\!
\left(\frac{\ln(\mu^2 {\Delta x}_{\scriptscriptstyle
+-}^2)}{{\Delta x}_{\scriptscriptstyle +-}^2} \! \right) \!-\!
\frac{H^2}{2^4 \pi^4} \frac{\ln\Bigl( \frac{\sqrt{e}}4 H^2 {\Delta
x}^2_{ \scriptscriptstyle +-} \Bigr)}{a a' {\Delta
x}^2_{\scriptscriptstyle +-}} \!+\! \frac{H^4}{2^6 \pi^4}
\ln^2\Bigl( \frac{\sqrt{e}}4 H^2 {\Delta x}^2_{
\scriptscriptstyle +-} \Bigr) , \qquad \\
\Bigl(i \Delta_{\scriptscriptstyle -+} \Bigr)^2 \!\!\!\! &
\rightarrow & \!\! - \frac{(a a')^{-2}}{2^6 \pi^4} \partial^2 \!\!
\left(\frac{\ln(\mu^2 {\Delta x}_{\scriptscriptstyle
-+}^2)}{{\Delta x}_{\scriptscriptstyle -+}^2} \! \right) \!-\!
\frac{H^2}{2^4 \pi^4} \frac{\ln\Bigl( \frac{\sqrt{e}}4 H^2 {\Delta
x}^2_{ \scriptscriptstyle -+} \Bigr)}{a a' {\Delta
x}^2_{\scriptscriptstyle -+}} \!+\! \frac{H^4}{2^6 \pi^4}
\ln^2\Bigl( \frac{\sqrt{e}}4 H^2 {\Delta x}^2_{ \scriptscriptstyle
-+} \Bigr) . \qquad
\end{eqnarray}
The square of the ${\scriptscriptstyle ++}$ propagator includes a
local divergent term in addition to the corresponding
${\scriptscriptstyle ++}$ finite terms,
\begin{eqnarray}
\Bigl(i \Delta_{\scriptscriptstyle ++}(x;x') \Bigr)^2 & \longrightarrow &
- \frac{i \mu^{-\epsilon}}{2^3 \pi^{2 \!-\! \frac{\epsilon}2}} \frac{\Gamma(1
\!-\! \frac{\epsilon}2)}{(1 \!-\! \epsilon) \epsilon} \, a^{-4 \!+\! 2\epsilon}
\delta^D(x\!-\! x') - \frac{(a a')^{-2}}{2^6 \pi^4} \partial^2 \left(
\frac{\ln(\mu^2 {\Delta x}_{\scriptscriptstyle ++}^2)}{{\Delta x}_{
\scriptscriptstyle ++}^2} \right) \nonumber \\
& & \hspace{2.5cm} - \frac{H^2}{2^4 \pi^4} \frac{\ln\Bigl( \frac{\sqrt{e}}4 H^2
{\Delta x}^2_{\scriptscriptstyle ++} \Bigr)}{a a' {\Delta x}^2_{
\scriptscriptstyle ++}} + \frac{H^4}{2^6 \pi^4} \ln^2\Bigl( \frac{\sqrt{e}}4
H^2 {\Delta x}^2_{\scriptscriptstyle ++} \Bigr) . \qquad \label{D2++}
\end{eqnarray}
The square of the ${\scriptscriptstyle --}$ propagator is the complex
conjugate,
\begin{eqnarray}
\Bigl(i \Delta_{\scriptscriptstyle --}(x;x') \Bigr)^2 &
\longrightarrow & \frac{i \mu^{-\epsilon}}{2^3 \pi^{2 \!-\!
\frac{\epsilon}2}} \frac{\Gamma(1 \!-\! \frac{\epsilon}2)}{(1
\!-\! \epsilon) \epsilon} \, a^{-4 \!+\! 2\epsilon}
\delta^D(x\!-\! x') - \frac{(a a')^{-2}}{2^6 \pi^4} \partial^2
\left( \frac{\ln(\mu^2 {\Delta x}_{\scriptscriptstyle
--}^2)}{{\Delta x}_{
\scriptscriptstyle --}^2} \right) \nonumber \\
& & \hspace{2.5cm} - \frac{H^2}{2^4 \pi^4} \frac{\ln\Bigl(
\frac{\sqrt{e}}4 H^2 {\Delta x}^2_{\scriptscriptstyle --}
\Bigr)}{a a' {\Delta x}^2_{ \scriptscriptstyle --}} +
\frac{H^4}{2^6 \pi^4} \ln^2\Bigl( \frac{\sqrt{e}}4 H^2 {\Delta
x}^2_{\scriptscriptstyle --} \Bigr) \; . \label{D2--}\qquad
\end{eqnarray}

\subsection{The Cube of a Propagator}

When the cube of a propagator multiplies a regular expression we again
decompose it according to (\ref{split}) but now we must retain $\epsilon
\neq 0$ in the $A^3$ and $3 A^2 B$ terms,
\begin{eqnarray}
\lefteqn{\Bigl(i \Delta(x;x') \Bigr)^3 = A^3 + 3 A^2 (B \!+\! C) +
3 A (B \!+\! C)^2 + (B \!+\! C)^3 , } \\
& & \longrightarrow \frac{\Gamma^3(1 \!-\! \frac{\epsilon}2)}{2^6 \pi^{6 -
\frac32 \epsilon}} \frac{(a a')^{-3 + \frac32 \epsilon}}{{\Delta x}^{6 - 3
\epsilon}} + \frac{3 H^{2-\e} \Gamma(1 \!-\! \frac{\epsilon}2) \Gamma(2 \!-\!
\epsilon)}{2^{7-\epsilon} \pi^{6 - \frac32 \epsilon} (a a')^{2 - \epsilon} }
\Bigg\{\!\!-\!\frac{2\Gamma(3 \!-\! \frac{\epsilon}2)}{\epsilon} \frac{\Gamma(2
\!-\!\frac{\epsilon}2)}{\Gamma(3 \!-\! \epsilon)} \, \frac{(\frac{H^2 a a'}4)^{
\frac{\epsilon}2}}{\Delta x^{4 - 3\epsilon}} \nonumber \\
& & \hspace{1.3cm} + \left[\frac{\pi \cot(\frac{\pi\epsilon}2) +
\ln(aa')}{ \Delta x^{4 - 2 \epsilon}} \right] \Bigg\} + \frac{3
H^4}{2^8 \pi^6} \frac{\ln^2\Bigl( \frac{\sqrt{e}}4 H^2 {\Delta
x}^2 \Bigr)}{a a' {\Delta x}^2} - \frac{H^6}{2^9 \pi^6}
\ln^3\Bigl( \frac{\sqrt{e}}4 H^2 {\Delta x}^2 \Bigr) \; . \qquad
\label{inmd}
\end{eqnarray}
The $A^3$ term is easy to read off from the identities
(\ref{++63}), (\ref{--63}) and (\ref{+-63}). For example, the
${\scriptscriptstyle +-}$ case gives,
\begin{equation}
\frac{\Gamma^3(1 \!-\! \frac{\epsilon}2)}{2^6 \pi^{6 - \frac32
\epsilon}} \frac{(a a')^{-3 + \frac32 \epsilon}}{{\Delta
x}_{\scriptscriptstyle +-}^{6 - 3 \epsilon}} \! \longrightarrow -
\frac{(a a')^{-3}}{2^{11} \pi^6} \,
\partial^4 \! \left[ \frac{\ln(\mu^2 \Delta x^2_{\scriptscriptstyle +-})}{
\Delta x^2_{\scriptscriptstyle +-}} \right] \! .\label{A^3}
\end{equation}
The $3 A^2 B$ term is more complicated on account of the
$1/\epsilon$ divergence in its prefactors. One must first
partially integrate and multiply before taking $\epsilon = 0$ in
the nonlocal term. For the $+-$ case we have, \be &&\lefteqn{3
(A^2B)_{+-} =\frac{3 H^{2-\e}(aa')^{-2+\e}\mu^{-2\e}}{2^{7-\e}
\pi^{6 - \frac{3}{2} \e}} \;
\G(1-\frac{\e}{2})}\nonumber\\
&&\hspace{0.7cm}\times\Bigg\{ \frac{(1 -
\frac{\epsilon}2)(1-\frac{\e}{4})\G^2(1-\frac{\e}{2})}{2(1-\frac{3}{2}\e)\e}\;
\Bigl(\frac{H^2 a a'}{4\mu^2}\Bigr)^{\frac{\epsilon}2}
\partial^2 \left[ \frac{\ln(\mu^2 \Delta x^2_{+-})}{\Delta x^2_{+-}}
+\e\frac{\ln^2(\mu^2 \Delta x^2_{+-})}{\Delta
x^2_{+-}}\right]\nonumber\\&&\hspace{0.7cm}-\frac{\G(1-\e)}{4}\left[\pi
\cot\left(\frac{\pi \epsilon}{2}\right)+\ln(aa')\right]
\;\partial^2 \left[ \frac{\ln(\mu^2 \Delta x^2_{+-})}{\Delta
x^2_{+-}} +\frac34\e\frac{\ln^2(\mu^2 \Delta x^2_{+-})}{\Delta
x^2_{+-}}\right] \Bigg\} . \label{A^2B} \ee At this point we
digress to recall some basic series expansions,
\begin{equation}
\Gamma(1-\epsilon) = 1 + \gamma \epsilon + \Bigl[\frac{\gamma^2}2
+ \frac{\pi^2}{12}\Bigr] \epsilon^2 + O(\epsilon^3) \qquad ,
\qquad \pi {\rm cot}\Bigl(\frac{\pi \epsilon}2\Bigr) =
\frac2{\epsilon} \left[1 - \frac{\pi^2}{12} \epsilon^2 +
O(\epsilon^4)\right] .\label{master}
\end{equation}
It follows that the various prefactors in (\ref{A^2B}) are,
\begin{eqnarray}
\frac{(1 \!-\! \frac{\epsilon}2) (1 \!-\!
\frac{\epsilon}4)\Gamma^2(1 \!-\! \frac{\epsilon}2)}{2 (1 \!-\!
\frac32 \epsilon) \epsilon} & = & \frac1{2 \epsilon} \!+\!
\frac{\g}{2} \!+\! \frac38\!+\! O(\epsilon)
\; , \\
- \frac{\G(1-\e)}{4}\pi \cot\Bigl(\frac{\pi\epsilon}2\Bigr) & = &
-\frac1{2\epsilon} \!-\! \frac\g{2}
\!+\! O(\epsilon) \; , \\
- \frac{\G(1-\e)}{4} & = & -\frac{1}{4}+ O(\epsilon) \; .
\end{eqnarray}
Making use of these expansions in (\ref{A^2B}) gives, \be
3(A^2B)_{+-}\longrightarrow\frac{3H^2(aa')^{-2}}{2^8\pi^6}
\partial^2\left[\ln\Big(\frac{He^{\frac{3}{4}}}{2\mu}\Big)
\frac{\ln(\mu^2 \Delta x^2_{+-})}{\Delta x^2_{+-}}
+\frac{1}{4}\frac{\ln^2(\mu^2 \Delta x^2_{+-})}{\Delta
x^2_{+-}}\right]\; .\label{A^2B+-}\ee Employing (\ref{A^3}) and
(\ref{A^2B+-}) in (\ref{inmd}) gives\be &&(i\D_{+-}(x ;
x'))^3\rightarrow - \frac{(a a')^{-3}}{2^{11} \pi^6} \,
\partial^4 \! \left[ \frac{\ln(\mu^2 \Delta x^2_{\scriptscriptstyle +-})}{
\Delta x^2_{\scriptscriptstyle +-}} \right]+
\frac{3H^2(aa')^{-2}}{2^8\pi^6}
\partial^2\Bigg[\ln\Big(\frac{He^{\frac{3}{4}}}{2\mu}\Big)
\frac{\ln(\mu^2 \Delta x^2_{+-})}{\Delta x^2_{+-}}\nonumber\\
&&\hspace{2cm}+\frac{1}{4}\frac{\ln^2(\mu^2 \Delta
x^2_{+-})}{\Delta x^2_{+-}}\Bigg]+\frac{3 H^4}{2^8 \pi^6}
\frac{\ln^2\Bigl( \frac{\sqrt{e}}4 H^2 \Delta x_{+-}^2 \Bigr)}{a
a' {\Delta x^2_{+-}}} - \frac{H^6}{2^9 \pi^6} \ln^3\Bigl(
\frac{\sqrt{e}}4 H^2 \Delta x^2_{+-} \Bigr)\; .\label{D3+-}\ee The
${\scriptscriptstyle -+}$ case follows by complex conjugation, \be
& & (i\D_{-+}(x ; x'))^3\rightarrow - \frac{(a a')^{-3}}{2^{11}
\pi^6} \,
\partial^4 \! \left[ \frac{\ln(\mu^2 \Delta x^2_{\scriptscriptstyle -+})}{
\Delta x^2_{\scriptscriptstyle -+}} \right]+
\frac{3H^2(aa')^{-2}}{2^8\pi^6}
\partial^2\Bigg[\ln\Big(\frac{He^{\frac{3}{4}}}{2\mu}\Big)
\frac{\ln(\mu^2 \Delta x^2_{-+})}{\Delta x^2_{-+}}\nonumber\\
&&\hspace{2cm}+\frac{1}{4}\frac{\ln^2(\mu^2 \Delta
x^2_{-+})}{\Delta x^2_{-+}}\Bigg]+\frac{3 H^4}{2^8 \pi^6}
\frac{\ln^2\Bigl( \frac{\sqrt{e}}4 H^2 \Delta x_{-+}^2 \Bigr)}{a
a' {\Delta x^2_{-+}}} - \frac{H^6}{2^9 \pi^6} \ln^3\Bigl(
\frac{\sqrt{e}}4 H^2 \Delta x^2_{-+} \Bigr)\; .\label{D3-+} \ee
The cube of the ${\scriptscriptstyle ++}$ propagator includes a
local divergent term in addition to the corresponding
${\scriptscriptstyle ++}$ finite terms, \be &&(i\D_{++}(x ;
x'))^3\rightarrow -\frac{i\mu^{-2\e} \Gamma^2(1-\frac{\e}2)}{2^9\;
\pi^{4-\e}} \; \frac{(a a')^{-3+ \frac{3}{2}\e}
\partial^2 \delta^D(x-x')}{ (1 \!-\! \frac{3}{2}\e ) (1 \!-\! \e)(1
\!-\!\frac{3}{4} \e)
\e}+\frac{i3H^{2-\e}\mu^{-\e}a^{-4+2\e}}{2^{6-\e} \pi^{4-\e}}
\;\Bigg\{ \nonumber \\
&& \frac{(1 - \frac{\epsilon}2)(1-\frac{\e}{4})\G^2(1-\frac{\e}{2})}{(1-\frac32
\e)\e^2}\; \Bigl(\frac{H a}{2\mu}\Bigr)^{\epsilon}-\frac{\G(1-\e)}{\e}\left[\pi
\cot\left(\frac{\pi \epsilon}{2}\right)+2\ln{(a)}\right]
\Bigg\}\d^D(x-x') \nonumber \\
&& - \frac{(a a')^{-3}}{2^{11} \pi^6} \,
\partial^4 \! \left[ \frac{\ln(\mu^2 \Delta x^2_{\scriptscriptstyle ++})}{
\Delta x^2_{\scriptscriptstyle ++}} \right]+
\frac{3H^2(aa')^{-2}}{2^8\pi^6}
\partial^2\Bigg[\ln\Big(\frac{He^{\frac{3}{4}}}{2\mu}\Big)
\frac{\ln(\mu^2 \Delta x^2_{++})}{\Delta
x^2_{++}}+\frac{1}{4}\frac{\ln^2(\mu^2 \Delta x^2_{++})}{\Delta
x^2_{++}}\Bigg]\nonumber\\&&\hspace{5.5cm}+\frac{3 H^4}{2^8 \pi^6}
\frac{\ln^2\Bigl( \frac{\sqrt{e}}4 H^2 \Delta x_{++}^2 \Bigr)}{a
a' {\Delta x^2_{++}}} - \frac{H^6}{2^9 \pi^6} \ln^3\Bigl(
\frac{\sqrt{e}}4 H^2 \Delta x^2_{++} \Bigr)\; .\label{D3++} \ee
The cube of the ${\scriptscriptstyle --}$ propagator is the
complex conjugate, \be && (i\D_{--}(x ; x'))^3\rightarrow
\frac{i\mu^{-2\e} \Gamma^2(1-\frac{\e}2)}{2^9\; \pi^{4-\e}} \;
\frac{(a a')^{-3+ \frac{3}{2}\e}
\partial^2 \delta^D(x-x')}{ (1 \!-\! \frac{3}{2}\e ) (1 \!-\! \e)(1
\!-\!\frac{3}{4} \e)
\e}-\frac{i3H^{2-\e}\mu^{-\e}a^{-4+2\e}}{2^{6-\e} \pi^{4-\e}}
\;\Bigg\{ \nonumber \\
&& \frac{(1 - \frac{\epsilon}2)(1-\frac{\e}{4})\G^2(1-\frac{\e}{2})}{(1-\frac32
\e)\e^2}\; \Bigl(\frac{H a}{2\mu}\Bigr)^{\epsilon}-\frac{\G(1-\e)}{\e}\left[\pi
\cot\left(\frac{\pi \epsilon}{2}\right)+2\ln{(a)}\right]
\Bigg\}\d^D(x-x')\nonumber\\&& - \frac{(a a')^{-3}}{2^{11} \pi^6}
\,
\partial^4 \! \left[ \frac{\ln(\mu^2 \Delta x^2_{\scriptscriptstyle --})}{
\Delta x^2_{\scriptscriptstyle --}} \right]+
\frac{3H^2(aa')^{-2}}{2^8\pi^6}
\partial^2\Bigg[\ln\Big(\frac{He^{\frac{3}{4}}}{2\mu}\Big)
\frac{\ln(\mu^2 \Delta x^2_{--})}{\Delta
x^2_{--}}+\frac{1}{4}\frac{\ln^2(\mu^2 \Delta x^2_{--})}{\Delta
x^2_{--}}\Bigg]\nonumber\\&&\hspace{5.5cm}+\frac{3 H^4}{2^8 \pi^6}
\frac{\ln^2\Bigl( \frac{\sqrt{e}}4 H^2 \Delta x_{--}^2 \Bigr)}{a
a' {\Delta x^2_{--}}} - \frac{H^6}{2^9 \pi^6} \ln^3\Bigl(
\frac{\sqrt{e}}4 H^2 \Delta x^2_{--} \Bigr)\; .\label{D3--}\ee

\subsection{Evaluating the Integrals for the Snowman Diagram}

The first integral we need in equation (\ref{smints}) is,
\be
I_1\equiv-\dd^2_0\int
d^4x'' a''^2\ln(a'')\left[\frac{\ln(\mu^2\D x^2_{++})}{\D
x^2_{++}}-\frac{\ln(\mu^2\D x^2_{+-})}{\D x^2_{+-}}\right] \; .
\ee
The first step is to make use of the identity,
\be
\frac{ \ln\left(\mu^2 \D x^2 \right)}{\D x^2} = \frac{\dd^2}{8} \left[
\ln^2\left(\mu^2 \D x^2\right) - 2 \ln\left(\mu^2 \D x^2\right)\right]
\label{ld} \; .
\ee
Because the derivative is with respect to $x^{\mu}$ we can move it outside
the integration over $x^{\prime\prime \mu}$. The remaining integrand possesses
only logarithmic singularities.

We define the coordinate separations,
\be
\D\eta \equiv \eta - \eta'' \qquad ,
\qquad r \equiv \Vert \vec{x} - \vec{x}'' \Vert \; . \label{id3}
\ee
The ${\scriptscriptstyle ++}$ and ${\scriptstyle +-}$ terms cancel for
$\Delta \eta < 0$ so we can restrict the integration to $\Delta \eta > 0$.
In this case the logarithms can be expanded as,
\begin{equation}
\ln\left[ \mu^2\D x^2_{\scriptscriptstyle +\pm} \right] = \ln\left[ \mu^2
(\D\eta^2 - r^2) \right] \pm i\pi \theta(\D \eta^2 - r^2) \; , \label{id4}
\end{equation}
We then perform the angular and radial integrations and act the first
three derivatives,
\be
I_1 &=& i2 \pi^2 \dd_0^4 \int_{\eta_i}^\eta d\eta''
a''^2 \ln\left( a'' \right)\int_0^{\D\eta} dr r^2 \left(\ln\left[
\mu^2 (\D \eta^2
- r^2)\right] - 1 \right) \nonumber \\
&=& i2 \pi^2 \dd_0^4 \int_{\eta_i}^\eta d\eta'' a''^2 \ln\left(a''
\right)\D \eta^3 \left[ {2 \over 3} \ln\left(2 \mu \D\eta
\right) - {11 \over 9} \right] \nonumber \\
&=& i8 \pi^2 \dd_0 \int_{\eta_i}^\eta d\eta'' a''^2 \ln\left(a''
\right)\ln\left(2 \mu \D\eta \right)\; .
\ee
The lower limit $\eta_i = -H^{-1}$ ($t'' = 0$) derives from the fact that
the state is released in free Bunch-Davies vacuum at this instant.

To perform the remaining temporal integration first change variables from
$\eta''$ to $a'' = -1/(H\eta'')$ and make the corresponding change in
the derivative,
\begin{equation}
I_1 = i 8 \pi^2 a^2 \frac{\partial}{\partial a} \int_1^a da'' \ln(a'')
\ln\left[ \frac{2 \mu}{H} \Bigl(\frac1{a''} - \frac1{a}\Bigr)\right] .
\end{equation}
Then break up the second logarithm and act the derivative on the
two nonsingular terms, \be I_1 & = & -i8
\pi^2a^2\left[\ln^2(a)-\ln\left(\frac{2\mu}{H}\right)\ln(a)-\frac{\dd}{\dd
a}\int_1^a da''\ln(a'')\ln\left(1-\frac{a''}{a}\right)\right]\; .
\label{137} \ee Now substitute the expansion, \beeq
\ln\left(1-\frac{a''}{a}\right)=-\sum_{n=1}^\infty
\frac{1}{n}\left(\frac{a''}{a}\right)^n\; ,\label{ex} \eneq and
integrate termwise, \be \int_1^a
da''\ln(a'')\ln\left(1-\frac{a''}{a}\right)=-a\Bigg\{
\ln(a)-2+\frac{\pi^2}{6}+\sum_{n=1}^\infty
\frac{a^{-(n+1)}}{n(n+1)^2}\Bigg\}\label{lnln}\; . \ee Combining
terms gives the following final result, \be I_1 & = & -i8 \pi^2
a^2 \left[\ln^2(a) - \ln\left(\frac{2\mu}{H}\right) \ln(a) +
\ln(a) - 1 + \frac{\pi^2}{6} - \sum_{n=1}^\infty
\frac{a^{-(n+1)}}{(n+1)^2} \right] \; . \label{I1} \ee

The second integral in (\ref{smints}) is,
\be
I_2\equiv\int d^4x''
a''^3\ln(a'')\Bigg[\frac{\ln(\frac{H^2\sqrt{e}}{4}\D x^2_{++})}{\D
x^2_{++}}-\frac{\ln(\frac{H^2\sqrt{e}}{4}\D x^2_{+-})}{\D
x^2_{+-}}\Bigg]\; .
\ee
Proceeding as before we extract a derivative with identity (\ref{ld})
and then exploit (\ref{id4}) to obtain,
\be
I_2&=&i\frac{\pi}{2}\dd^2\int
d^4x'' a''^3\ln(a'') \theta(\D\eta^2-r^2)
\ln\left(\frac{H^2}{4\sqrt{e}}(\D\eta^2-r^2)\right) \;
\nonumber\\
&=&i2\pi^2(-\dd^2_0)\int_{\eta_i}^\eta d\eta''
a''^3\ln(a'')\int_0^{\D\eta} dr r^2
\ln\left(\frac{H^2}{4\sqrt{e}}(\D\eta^2-r^2)\right) \; .
\ee
Now perform the angular integrations, make the change of variable
$r\equiv \D\eta\; z$, act the temporal derivatives and change
variables from $\eta''$ to $a''$,
\be \!\!\!\!\!\!\!\!\!\!\!\!\!\!\!\!\!\!\!\!\!\!\!\!\!\!\!\!\!
I_2 \!\! & = & \!\!-i2\pi^2\!\!\int_{\eta_i}^\eta
d\eta'' a''^3\ln(a'')\int_0^1 dz z^2\dd_0^2\left[
\D\eta^3\ln\left(\frac{H^2}{4\sqrt{e}}\D\eta^2(1-z^2)\right)\right]\nonumber \\
\!\! & = & \!\! -i2 \pi^2 \!\! \int_{\eta_i}^\eta d\eta'' a''^3 \ln(a'')
\D\eta \int_0^1 dz z^2\left\{6 \ln\left[ \frac{H^2}{4} \D\eta^2 (1-z^2)
\right] + 7 \right\} \nonumber \\
\!\! & = & \!\! - \frac{i 2 \pi^2}{H^2} \!\! \int_1^a da''\ln(a'') \left(1
- \frac{a''}{a} \right) \!\! \left\{\! 4 \ln\left(\frac{1}{a''} -
\frac{1}{a}\right)\! - 3 \right\}\; . \label{1-z}
\ee
A small rearrangement gives,
\be
\!\!\!\!\!\!\!\!\!\!\!\!\!\!\!\!\!\! I_2 \!\! & = & \!\! \frac{i 2 \pi^2}{H^2}
\Bigg\{\! \int_1^a \!\!\! da'' \!\left( \!1-\! \frac{a''}{a} \right)\!
\left[3 \ln(a'') + 4 \ln^2(a'') \right] \!-4\! \int_1^a\! da'' \!\left(\!1
\!-\! \frac{a''}{a} \right)\! \ln(a'') \ln\!\left(\!1 \!-\! \frac{a''}{a}
\right)\!\! \Bigg\} \; . \label{second}
\ee
The only non-trivial integral left in (\ref{second}) can be evaluated using
(\ref{ex})
\be
\!\!\!\!\!\!\!\!\int_1^a \!\!da''
\!\!\left(1-\frac{a''}{a}\right)\!\ln(a'')\ln\!\!\left(\!1-\frac{a''}{a}\right)\!=\!-a\Bigg\{
\frac{1}{4}\ln(a)\!-\!1\!+\!\frac{\pi^2}{12}\!+\!\!\sum_{n=1}^\infty\!\left[\!
\frac{a^{-(n+1)}}{n(n+1)^2}\!-\!\frac{a^{-(n+2)}}{n(n+2)^2}\!\right]
\!\!\Bigg\}\; .
\ee
The final result for $I_2$ is therefore,
\be
\!\!\!\!\!\!\!\! I_2\!\! & = & \!\! \frac{i 2 \pi^2 a}{H^2}
\Bigg\{2\ln^2(a)\!-\!\frac{7}{2}\ln(a)\!+\!\frac{3}{4}
\!+\!\frac{\pi^2}{3}\!-\!5a^{-1}\!+\!\frac{1}{4}a^{-2}\!+\!4\!\sum_{n=1}^\infty
\!\left[\!\frac{a^{-(n+1)}}{n(n+1)^2} \!-\! \frac{a^{-(n+2)}}{n(n+2)^2}\!
\right]\!\Bigg\}\label{I2}\; .
\ee

The initial reduction of the last integral in ({\ref{smints}}) is familiar
from $I_1$ and $I_2$,
\be
& & \!\!\!\!\!\! I_3 \equiv \int d^4x'' a''^4 \ln(a'') \Bigg[ \ln^2 \left(
\frac{H^2\sqrt{e}}{4} \D x^2_{\scriptscriptstyle ++}\right) - \ln^2\left(
\frac{H^2\sqrt{e}}{4}\D x^2_{\scriptscriptstyle +-}\right)\Bigg] \nonumber \\
& & = i 4 \pi \int d^4x'' a''^4 \ln(a'') \theta(\D\eta^2 - r^2) \ln\left(
\frac{H^2\sqrt{e}}{4} (\D\eta^2 - r^2) \right) \; \nonumber \\
& & = i 16 \pi^2 \int^1_0 dz z^2 \int_{\eta_i}^\eta d\eta'' a''^4 \ln(a'')
\D\eta^3 \ln\left( \frac{H^2\sqrt{e}}{4} \D\eta^2 (1-z^2) \right) \nonumber \\
& & = \frac{i 16\pi^2}{H^4} \int^1_0 dz z^2 \int_{1}^a \frac{da''}{a''}
\ln(a'') \left(1 - \frac{a''}{a} \right)^3 \ln\left(\frac{\sqrt{e}}{4}
\left(\frac{1}{a''} - \frac{1}{a} \right)^2 (1-z^2) \right) \nonumber \\
& & = \frac{i 16\pi^2}{H^4} \int_{1}^a \frac{da''}{a''} \ln(a'') \left(1 -
\frac{a''}{a} \right)^3 \left\{ \frac23 \ln\Bigl(\frac1{a''} - \frac1{a}\Bigr)
- \frac{13}{18} \right\} \; .
\ee
At this stage it is best to integrate by parts on the term $\ln(a'')/a''
= \partial/{\partial a''} \Bigl(\ln^2(a'')/2\Bigr)$,
\be
& & \!\!\!\!\!\! I_3 = \frac{i 16 \pi^2}{H^4} \Bigg\{- \int_1^a da''
\left(1 - \frac{a''}{a} \right)^2 \ln^2(a'') \left[ \frac{1}{a} \left(
\frac{13}{12} + \ln(a'') \right) - \frac{1}{3a''} \right] \nonumber \\
& & \hspace{5.5cm} + \frac{1}{a} \int_1^a da'' \left(1 - \frac{a''}{a}
\right)^2 \ln^2(a'') \ln\left(1 - \frac{a''}{a} \right) \Bigg\} \; .\label{ln2}
\ee
The first term is straightforward and the final term can be evaluated using
(\ref{ex}),
\be
& & \int_1^a da'' \left(1 - \frac{a''}{a} \right)^2 \ln^2(a'') \ln\left(1
- \frac{a''}{a} \right) = a \Bigg\{ -\frac{80}{27} + \frac{11}{54} \pi^2
+ \frac23 \zeta(3) + \left(\frac{71}{54} - \frac{\pi^2}{9} \right) \ln(a)
\nonumber \\
& & \hspace{4cm} - \frac19 \ln^2(a) + 2 \sum_{n=1}^\infty \left[ \frac{a^{-
(n+1)}}{n(n+1)^3} - \frac{2a^{-(n+2)}}{n(n+2)^3} + \frac{a^{-(n+3)}}{n(n+3)^3}
\right] \Bigg\} \; .\label{int0}
\ee
Substituting (\ref{int0}) in (\ref{ln2}) and adding everything up gives,
\be
&&I_3=i\frac{8\pi^2}{H^4}\Bigg\{-\frac{385}{324}+\frac{11}{27}\pi^2+\frac{4}{3}\zeta(3)
+\frac{a^{-3}}{81}+\frac{a^{-2}}{4}-5a^{-1}-\left(\frac{11}{6}
+\frac{2\pi^2}{9}\right)\ln(a)+\frac{31}{18}\ln^2(a)\nonumber\\&&\hspace{4cm}-\frac{4}{9}\ln^3(a)+4\sum_{n=1}^\infty\Bigg[
\frac{a^{-(n+1)}}{n(n+1)^3}-\frac{2a^{-(n+2)}}{n(n+2)^3}+\frac{a^{-(n+3)}}{n(n+3)^3}\Bigg]
\Bigg\}\label{I3}\; .
\ee

\centerline{\bf Acknowledgments}

This work was partially supported by NSF grant PHY-0244714
and by the Institute for Fundamental Theory.

\end{document}